\documentclass[12pt]{article}
\usepackage[pdftex]{graphicx}

\usepackage{amsmath,amsfonts}
\usepackage{amssymb}
\usepackage{amsthm}
\usepackage{times,txfonts}

\usepackage{xcolor} 

\usepackage{authblk}

\def\ul{\underline}

\usepackage{graphicx,color}
\usepackage{pict2e}

\newtheorem{thm}{Theorem}[section]
\newtheorem{prop}[thm]{Proposition}
\newtheorem{cor}[thm]{Corollary}
\newtheorem{lemma}[thm]{Lemma}
\newtheorem{dfn}[thm]{Definition}

\newtheorem{remark}[thm]{\it Remark}
\newtheorem{example}[thm]{\it Example}

\numberwithin{equation}{section}

\def\pf{\noindent{\it Proof.} \ }

\def\qed{\hfill $\square$}

\begin{document}

\title{Birational Weyl group actions and $q$-Painlev\'e equations via mutation combinatorics in cluster algebras}

\author{Tetsu Masuda, Naoto Okubo and Teruhisa Tsuda}
\date{}

\normalsize

\maketitle
\begin{abstract}
A cluster algebra is an algebraic structure generated by operations of a 
quiver (a directed graph) called the mutations and their
associated simple birational mappings.
By using a graph-combinatorial approach, 
 we present a systematic way to derive a 
tropical, i.e. subtraction-free birational, 
representation of Weyl groups from cluster algebras.
Our results provide 
an extensive class 
of Weyl group actions,   
including previously known examples with algebro-geometric background,
and hence are relevant to 
the $q$-Painlev\'e equations and their higher-order extensions.
Key ingredients of the argument are the combinatorial aspects of 
the reflection 
associated with a cycle subgraph in the quiver.
We also study symplectic structures of the discrete dynamical systems thus obtained.
The normal form of a skew-symmetric integer matrix
allows us to choose Darboux coordinates while preserving the birationality.
\end{abstract}

\renewcommand{\thefootnote}{\fnsymbol{footnote}}
\footnotetext{{\it 2020 Mathematics Subject Classification} 
13F60, 
14E07, 
20F55, 
34M55, 
 39A13. 
}
\footnotetext{{\it Keywords}: cluster algebra, birational Weyl group action, discrete dynamical system, mutation combinatorics, $q$-Painlev\'e equation.}


\tableofcontents

\section{Introduction}  \label{sect:intro}
It is classically known that an algebro-geometric setup is effective in
constructing birational representations of Weyl groups.
The configuration space of $n$ points in general position in the projective
space ${\mathbb P}^{m-1}$ naturally
possesses a birational action of the Weyl group corresponding to the T-shaped Dynkin diagram (see e.g. \cite{DO, KMNOY, Takenawa} and references therein):
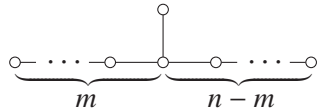
\begin{center}
\begin{picture}(160,45) 
\put(80,20){\circle{4}}  
\put(80,22){\line(0,1){16}}
\put(80,40){\circle{4}} 

\put(78,20){\line(-1,0){16}}
\put(60,20){\circle{4}}  
\put(58,20){\line(-1,0){6}}
\put(36,17){$\cdots$}
\put(32,20){\line(-1,0){6}}
\put(24,20){\circle{4}}  

\put(82,20){\line(1,0){16}}
\put(100,20){\circle{4}}  
\put(102,20){\line(1,0){6}}
\put(112,17){$\cdots$}
\put(128,20){\line(1,0){6}}
\put(136,20){\circle{4}}  
\put(24,17){\footnotesize$\underbrace{\qquad \qquad \quad \ \ }$}
\put(47,3){\small$m$}
\put(81,17){\footnotesize$\underbrace{\qquad \qquad \quad \ \ }$}
\put(97,3){\small$n-m$}
\end{picture}
\end{center}
In particular, if $(m,n)=(3,9)$ then the affine Weyl group of type $E^{(1)}_8$ occurs and its lattice part 
gives rise to 
the elliptic-difference Painlev\'e equation
\cite{ORG, Sakai1}.
 This case was explored by Sakai \cite{Sakai1}
 to clarify the 
 geometric origin 
 of the affine Weyl group symmetry of the Painlev\'e equations; 
 he classified all degenerations of the nine-point configuration in ${\mathbb P}^2$ and completed the list of (second-order) 
 discrete 
 Painlev\'e equations.
 Besides, even in the two-dimensional case, some special configurations of point sets that are not only nine points lead to various Dynkin diagrams \cite{Tsuda1}.
Furthermore,
in the higher-dimensional case, 
a geometric approach to birational representations of Weyl groups 
was proposed in \cite{TT}
 by means of pseudo-isomorphisms of a certain rational variety blown-up from $({\mathbb P}^1)^N$ along subvarieties
for the comb-shaped Dynkin diagram $T^{\boldsymbol k}_{\boldsymbol \ell}$:
\begin{center}
\begin{picture}(200,125)
\put(98,0){\small$n$}
\put(130,0){$n+1$}
\put(48,0){\small$n-1$}
\put(167,0){$\cdots$}
\put(20,0){$\cdots$}

\put(100,70){\circle{4}}
\put(102,70){\line(1,0){36}}  
\put(140,70){\circle{4}}
\put(142,70){\line(1,0){20}} 
\put(167,67){$\cdots$}

\put(98,70){\line(-1,0){36}}  
\put(60,70){\circle{4}} 
\put(58,70){\line(-1,0){20}}  
\put(20,67){$\cdots$}

\put(90,94){\footnotesize$\left\{\begin{array}{l}  \\ \\ \\ \\ \end{array} \right.$}
\put(78,94){\small$k_n$}
\put(100,72){\line(0,1){10}}  
\put(100,84){\circle{4}} 
\put(100,86){\line(0,1){6}} 
\put(98.5,96){$\vdots$}
\put(100,116){\line(0,-1){6}} 
\put(100,118){\circle{4}}
\put(90,41){\footnotesize$\left\{\begin{array}{l}  \\ \\ \\ \\ \end{array} \right.$}
\put(78,41){\small$\ell_n$}
\put(100,68){\line(0,-1){10}}  
\put(100,56){\circle{4}} 
\put(100,54){\line(0,-1){6}} 
\put(98.5,35){$\vdots$}
\put(100,24){\line(0,1){6}} 
\put(100,22){\circle{4}}

\put(50,94){\footnotesize$\left\{\begin{array}{l}  \\ \\ \\ \\ \end{array} \right.$}
\put(30,94){\small$k_{n-1}$}
\put(60,72){\line(0,1){10}}  
\put(60,84){\circle{4}} 
\put(60,86){\line(0,1){6}} 
\put(58.5,96){$\vdots$}
\put(60,116){\line(0,-1){6}} 
\put(60,118){\circle{4}}
\put(50,41){\footnotesize$\left\{\begin{array}{l}  \\ \\ \\ \\ \end{array} \right.$}
\put(30,41){\small$\ell_{n-1}$}
\put(60,68){\line(0,-1){10}}  
\put(60,56){\circle{4}} 
\put(60,54){\line(0,-1){6}} 
\put(58.5,35){$\vdots$}
\put(60,24){\line(0,1){6}} 
\put(60,22){\circle{4}}

\put(130,94){\footnotesize$\left\{\begin{array}{l}  \\ \\ \\ \\ \end{array} \right.$}
\put(110,94){\small$k_{n+1}$}
\put(140,72){\line(0,1){10}}  
\put(140,84){\circle{4}} 
\put(140,86){\line(0,1){6}} 
\put(138.5,96){$\vdots$}
\put(140,116){\line(0,-1){6}} 
\put(140,118){\circle{4}}

\put(130,41){\footnotesize$\left\{\begin{array}{l}  \\ \\ \\ \\ \end{array} \right.$}
\put(110,41){\small$\ell_{n+1}$}
\put(140,68){\line(0,-1){10}}  
\put(140,56){\circle{4}} 
\put(140,54){\line(0,-1){6}} 
\put(138.5,35){$\vdots$}
\put(140,24){\line(0,1){6}} 
\put(140,22){\circle{4}}

\end{picture}
\end{center}
specified by an arbitrary pair of sequences 
 ${\boldsymbol k}=(k_1,k_2,\ldots,k_N)$, ${\boldsymbol \ell}=(\ell_1,\ell_2,\ldots, \ell_N) \in ({\mathbb Z}_{>0})^N$.
This is thought of as an origin of 
higher-order $q$-Painlev\'e equations; 
see \cite{Masuda} in which
the higher-order $q$-Painlev\'e equation of type $D^{(1)}_{n}$
was derived.

On the other hand, it is known that some $q$-Painlev\'e equations can be described as 
birational mappings generated by sequences of mutations in cluster algebras associated with quivers appropriately chosen; see \cite{HI, Okubo1, Okubo2}.
A subsequent work by Bershtein--Gavrylenko--Marshakov \cite{BGM} (cf. \cite{Mizuno}) shows that all the (second-order)  $q$-Painlev\'e equations in Sakai's list \cite{Sakai1},
together with their underlying Weyl group symmetries, 
can be derived from cluster algebras 
in connection with
deautonomization of cluster integrable systems.

The aim of this paper is to
present a systematic way to derive 
birational 
representations of Weyl groups from cluster algebras by means of a graph-combinatorial point of view.
Key ingredients of the argument are the combinatorial aspects of the {\it reflection}
associated with a cycle subgraph in the quiver, 
which is defined by a certain sequence of mutations; see (\ref{eq:ref}).
Note that the reflection itself already appears in several areas of mathematics, including the higher-dimensional  Teichm\"uller theory; see \cite{GS, IIO} and also Remark~\ref{remark:history}.
After clarifying basic properties of the reflection,
we construct  birational Weyl group actions 
by assembling cycle graphs suitably. 
There are two main advantages of our usage of cluster algebras:
one is that the relations among reflections can be proved
by a simple combinatorial consideration 
with the aid of topological properties of the quivers.
The other is that  
the birational Weyl group actions are
derived 
without any algebro-geometric setup such as constructing the space of initial conditions.

Our framework includes the previously known examples of birational Weyl group actions such as \cite{KNY, Tsuda1, TT, Yamada} that are relevant to 
the $q$-Painlev\'e equations and their higher-order extensions.
If restricted to two-dimensional cases, 
the exceptions are only a few relatively degenerate ones in which some generators of the Weyl group are not associated with cycle subgraphs; 
see \cite{MOT} for details.
Obviously, our representation is {\it tropical}, i.e. given in terms of subtraction-free birational mappings \cite{Kirillov},
and 
hence
admits a combinatorial counterpart via the ultra-discretization \cite{TTMS}.
We also discuss symplectic structures of the discrete dynamical systems 
thus obtained.
By using the normal form of a skew-symmetric integer matrix, 
Darboux coordinates are obtained from the Poisson structure possessed by a cluster algebra  while preserving the birationality. 
\\

In the next section we begin by 
preparing 
some basic notions of cluster algebras. 
Our main interest is the birational action of a group consisting of compositions of mutations and permutations of vertices that keeps the quiver invariant.
In Section~\ref{sect:cycle}, 
we introduce the {\it reflection} associated with a cycle graph.
We determine the necessary and sufficient condition for a quiver containing a cycle subgraph to be invariant under the associated reflection (Theorem~\ref{thm:R-inv}).
An explicit formula for the birational action of the reflection (Proposition~\ref{prop:birat}) 
reveals its rotational symmetry (\ref{eq:rotation}) 
and, thereby, permutation symmetry
(Corollary~\ref{cor:sym}), which are crucial 
in the following discussion.
Section~\ref{sect:rel} concerns a quiver that contains two or more cycles and is invariant under the associated reflections.
We prove the relations satisfied by the reflections for some specific combinations of cycle subgraphs:  
two intersecting cycles (Theorem~\ref{thm:commute}),
 two cycles connected with a hinge (Theorem~\ref{thm:hinge})
 and 
 two adjacent cycles in a ladder shape  (Theorem~\ref{thm:ladder}).
 For instance, the reflections associated with two intersecting cycles turn out to be commutative.
Based on these results, 
we enjoy an extensive class of birational representations of Weyl groups,
which in the affine case yields
discrete dynamical systems of Painlev\'e type as lattice parts.
In Section~\ref{sect:ex}, 
we demonstrate the construction of Weyl group actions from cluster algebras through typical examples related to 
 the $q$-Painlev\'e equations.
 Section~\ref{sect:symp} is devoted to symplectic structures of the discrete dynamical systems arising from cluster algebras.
A unified way to choose Darboux coordinates is presented.

\begin{remark}\rm
It is known that
the $q$-Painlev\'e equations, or their underlying Weyl group symmetries, 
admit {\it $\tau$-function formalism} in which dynamical variables possess a certain regularity analogous to the Laurent phenomenon;
see e.g. \cite{Tsuda1, TT}. 
In a sequel \cite{MMOTT} to the present paper, 
we explore $\tau$-function formalism for our framework of 
birational Weyl group actions, 
by employing a non-normalized cluster algebra equipped with two series of variables.
\end{remark}

\section{Preliminaries from cluster algebras} 
\label{sect:prelim}

A cluster algebra is an algebraic structure generated by operations of a 
quiver, called the {\it mutations}, and their
associated simple birational mappings.
In this section we prepare some basic notions of cluster algebras required to present our results, according to Fomin--Zelevinsky \cite{FZ}.

Let $Q=(V,E)$ be a quiver, i.e. 
a directed graph, given by a set of vertices 
$V=\{1,2,\ldots,N\}$ and a set of edges $E \subseteq V \times V$.
Assume that $Q$ has no  loops $i \to i$ nor $2$-cycles $ i \to j \to i$ 
but may have multiple edges.
We identify $Q$ with a skew-symmetric integer matrix $B=(b_{ij})_{i,j=1}^N$,
called the {\it signed adjacency matrix} of $Q$, such that
\[
b_{ij}=-b_{ji}= \text{(the number of edges $i \to j$)} \quad \text{if} \quad b_{ij}>0.
\]
Let ${\boldsymbol y}=(y_1,y_2,\ldots,y_N)$ be an $N$-tuple of algebraically independent and commutative variables.
The pair $(Q,{\boldsymbol y})$ is called an  ({\it initial}) {\it Y-seed}.

We will define the ({\it seed}) {\it mutation}
$(Q', {\boldsymbol y'})=\mu_k(Q, {\boldsymbol y})$
in direction  $k \in V$.
The mutated quiver $Q'=\mu_k(Q)$ is obtained by the procedure below:
\begin{quote}
1. Add a new edge $i\to j$  for each subgraph $i\to k\to j$;
\\
2. reverse the orientation of all edges containing $k$;
\\ 
3. remove the $2$-cycles that appear.
\end{quote}
Alternatively, we may describe this procedure in terms of the signed adjacency matrix $B=(b_{ij})_{i,j=1}^N$ of $Q$ as 
\begin{equation}  \label{eq:mut_B}
B'=\mu_k(B)= {}^{\rm T} A_k B A_k,
\end{equation}
where
\begin{equation}
\label{eq:Ak}
A_k=
 \begin{pmatrix} 1 &
\\& \ddots
\\
&&1
\\
[b_{k1}]_+ & \cdots & [b_{k,k-1}]_+ & -1 & [b_{k,k+1}]_+ & \cdots & [b_{k,N}]_+
\\
&&&&1
\\
&&&&&\ddots
\\
&&&&&& 1
\end{pmatrix}
 \leftarrow  \text{the $k$th row}
\end{equation}
and $[a]_+=\max \{a,0\}$ for $a \in {\mathbb R}$.
I.e.,
\[
b_{ij}'=\left\{
\begin{array}{lll}
-b_{ij} &&\text{($i=k$ or $j=k$)}\\
b_{ij}+b_{ik}[b_{kj}]_+ +[-b_{ik}]_+ b_{kj}
&&\text{($i, j \ne k$)}.
\end{array}
\right.
\]
In parallel, the mutated $y$-variables
${\boldsymbol y'}=\mu_k({\boldsymbol y})$ 
are defined by the birational transformations
\begin{equation} \label{eq:mut_y}
y_i'=\left\{
\begin{array}{lll}
{y_k}^{-1}&&(i=k)\\
y_i  {y_k}^{ [b_{ki}]_+ } (1+y_k)^{-b_{ki}}
&&(i\ne k).
\end{array}
\right.
\end{equation}
Suppose a composition of mutations 
$w=\mu_{i_1} \circ\mu_{i_2} \circ \cdots \circ \mu_{i_\ell} $ 
acts on a rational function $\varphi=\varphi({\boldsymbol y})$ as
$w . \varphi({\boldsymbol y})=\varphi( {\boldsymbol y} . w)$,
i.e. $w$ acts on $y$-variables from the right.
Then it holds that
\[
{\mu_k}^2 ={\rm id} \quad \text{(involution)}
\quad \text{and} \quad
\mu_i \circ\mu_j =\mu_j \circ\mu_i \quad \text{if} \quad b_{ij}=0.
\]
A symmetric group  ${\frak S}_N$ acts on the quiver $Q$ as permutations of the labels of vertices, which is naturally extended to $y$-variables as
$\sigma(y_i)= y_{\sigma^{-1}(i)} $ $(\sigma \in {\frak S}_N)$.
Then it holds that
\[(i,j)\circ  \mu_i=\mu_j \circ (i,j) \]
for any  $i, j \in V$. 
Also, we consider an operation $\iota$,
called the {\it inversion},
which reverses the orientation of all edges of $Q$ with 
$\iota(y_i)={y_i}^{-1}$. 
We see that $\iota$ commutes with any mutation.

\begin{example}[Computation of a composition of mutations]
\rm
Apply the composition $w=\mu_2 \circ \mu_1$ of mutations to the Y-seed:
\[
Q=
\begin{array}{l}
\begin{picture}(50,50)
\put(0,40){$1$}
\put(0,0){$3$}
\put(40,40){$2$}
\put(9,42){\vector(1,0){26}}
\put(3,10){\vector(0,1){26}}
\put(36,36){\vector(-1,-1){26}}
\end{picture}
\end{array} \quad \text{and}
\quad {\boldsymbol y}=(y_1,y_2,y_3).
\]
The mutated quivers are obtained 
as follows:
\[
\mu_1(Q)=
\begin{array}{l}
\begin{picture}(50,50)
\put(0,40){$1$}
\put(0,0){$3$}
\put(40,40){$2$}
\put(35,42){\vector(-1,0){26}}
\put(3,36){\vector(0,-1){26}}
\end{picture}
\end{array},
\qquad 
\mu_2 \circ \mu_1(Q)=
\begin{array}{l}
\begin{picture}(50,50)
\put(0,40){$1$}
\put(0,0){$3$}
\put(40,40){$2$}
\put(9,42){\vector(1,0){26}}
\put(3,36){\vector(0,-1){26}}
\end{picture}
\end{array}
\]
On the other hand, the mutated $y$-variables
are determined as
\[
\mu_1(y_1)={y_1}^{-1}, \quad \mu_1(y_2)=y_2\left(1+{y_1}^{-1}\right)^{-1}, \quad
\mu_1(y_3)=y_3(1+y_1)
\]
at the quiver $Q$, and
\[
\mu_2(y_1)=y_1\left(1+{y_2}^{-1}\right)^{-1}, \quad
\mu_2(y_2)={y_2}^{-1}, 
\quad \mu_2(y_3)=y_3
\]
at the intermediate quiver $\mu_1(Q)$;
therefore, 
by the composition rule of mutations, we get
\begin{align*}
\mu_2 \circ \mu_1(y_1)&= \mu_1(y_1)\left(1+{\mu_1(y_2)}^{-1}\right)^{-1}=
{y_1}^{-1} \left(1+ {y_2}^{-1} \left(1+{y_1}^{-1}\right) \right)^{-1} 
= 
\frac{y_2}{1+y_1+y_1y_2}
, 
\\
\mu_2 \circ \mu_1(y_2)&={\mu_1(y_2)}^{-1}={y_2}^{-1}\left(1+{y_1}^{-1}\right)
=\frac{1+y_1}{y_1y_2}, 
\\
 \mu_2 \circ \mu_1(y_3)&=\mu_1(y_3)=y_3(1+y_1).
\end{align*}
\end{example}

Each mutation $\mu_k$, permutation $ \sigma \in {\frak S}_N$ and the inversion  $\iota$ generally changes the quiver $Q$.
Let $G_Q$ denote the whole set of compositions of mutations, permutations and the inversion that keep $Q$ invariant. Then $G_Q$ provides, via the above actions on $y$-variables, a nontrivial group of birational transformations on the field ${\mathbb Q}(y_1, y_2,\ldots,y_N)$ of rational functions; this is the subject we are interested in.

\section{Cycle graphs and reflections}
\label{sect:cycle}
In this section we introduce the reflection associated with a cycle graph.
We determine the necessary and sufficient condition for a quiver containing a cycle subgraph to be invariant under the associated reflection.
An explicit formula for the birational action of the reflection reveals its rotational symmetry, which will be crucial to investigating relations between two reflections in Section~\ref{sect:rel}.

Let $n$ be an integer greater than one.
We use the notation
$\mu_{i_1,i_2, \ldots, i_\ell}=
\mu_{i_1} \circ \mu_{i_2} \circ \cdots \circ \mu_{i_\ell}$
for the sake of brevity. 
First we consider an oriented cycle of length $n$:
\[
C= (1 \to 2 \to \cdots \to n \to 1)=\left\{  
\begin{array}{l}
\begin{picture}(150,50)
\put(10,30){\circle{4}}  
\put(6,38){$1$} 
\put(12,30){\vector(1,0){16}}
\put(30,30){\circle{4}}  
\put(26,38){$2$} 
\put(32,30){\vector(1,0){16}}
\put(50,30){\circle{4}}  
\put(46,38){$3$} 
\put(52,30){\vector(1,0){36}}
\put(62, 38){$\cdots$}
\put(90,30){\circle{4}}  
\put(86,38){$n-1$} 

\put(50,10){\circle{4}} 
\put(46,-4){$n$} 
\put(48,11){\vector(-2,1){36}}  
\put(88,29){\vector(-2,-1){36}} 
\put(140,10){$(n \geq 3)$}
\end{picture}
\\
\begin{picture}(130,50)
\put(10,10){\circle{4}}  
\put(6,18){$1$} 
\put(30,10){\circle{4}}  
\put(26,18){$2$} 
\put(140,10){$(n =2)$}
\end{picture}
\end{array}
\right.
\]
Apply the composition $M=\mu_{n-1, \ldots,2,1}=\mu_{n-1} \circ \cdots \circ \mu_2 \circ \mu_{1}$
of mutations to $C$. We can chase the mutated quivers as follows:
\begin{center}
\begin{picture}(430,170)

\put(0,150){$C=$}
\put(40,150){\circle{4}}  
\put(36,158){$1$} 
\put(42,150){\vector(1,0){16}}
\put(60,150){\circle{4}}  
\put(56,158){$2$} 
\put(62,150){\vector(1,0){16}}
\put(80,150){\circle{4}}  
\put(76,158){$3$} 
\put(82,150){\vector(1,0){36}}
\put(92, 158){$\cdots$}
\put(120,150){\circle{4}}  
\put(116,158){$n-1$} 

\put(80,130){\circle{4}} 
\put(76,116){$n$} 
\put(78,131){\vector(-2,1){36}}  
\put(118,149){\vector(-2,-1){36}}  

\put(150,150){$\stackrel{\mu_1}{\longrightarrow} $}

\put(190,150){\circle{4}}  
\put(186,158){$1$} 
\put(208,150){\vector(-1,0){16}}
\put(210,150){\circle{4}}  
\put(206,158){$2$} 
\put(212,150){\vector(1,0){16}}
\put(230,150){\circle{4}}  
\put(226,158){$3$} 
\put(232,150){\vector(1,0){36}}
\put(242, 158){$\cdots$}
\put(270,150){\circle{4}}  
\put(266,158){$n-1$} 

\put(230,130){\circle{4}} 
\put(226,116){$n$} 
\put(192,149){\vector(2,-1){36}} 
\put(268,149){\vector(-2,-1){36}}  

\put(228,131){\vector(-1,1){17}}  

\put(300,150){$\stackrel{\mu_2}{\longrightarrow} $}

\put(340,150){\circle{4}}  
\put(336,158){$1$} 
\put(342,150){\vector(1,0){16}}
\put(360,150){\circle{4}}  
\put(356,158){$2$} 
\put(378,150){\vector(-1,0){16}}
\put(380,150){\circle{4}}  
\put(376,158){$3$} 
\put(382,150){\vector(1,0){36}}
\put(392, 158){$\cdots$}
\put(420,150){\circle{4}}  
\put(416,158){$n-1$} 

\put(380,130){\circle{4}} 
\put(376,116){$n$} 
\put(418,149){\vector(-2,-1){36}}  

\put(361,148){\vector(1,-1){17}}  

\put(380,132){\vector(0,1){16}}
\put(185,90){``A triangle $(n \to k+1 \to k \to n)$ shifts to the right."}
 
\put(10,40){$\cdots$ \quad $\stackrel{\mu_{n-2}}{\longrightarrow} $}

\put(80,40){\circle{4}}  
\put(76,48){$1$} 

\put(82,40){\vector(1,0){56}}

\put(92, 48){$\cdots$}

\put(140,40){\circle{4}}  
\put(120,48){$n-2$} 

\put(160,40){\circle{4}}  
\put(156,48){$n-1$} 

\put(158,40){\vector(-1,0){16}}

\put(140,20){\circle{4}} 
\put(136,6){$n$} 

\put(140,38){\vector(0,-1){16}} 

\put(190,40){$\stackrel{\mu_{n-1}}{\longrightarrow} $}

\put(230,40){\circle{4}}  
\put(226,48){$1$} 

\put(232,40){\vector(1,0){56}}

\put(242, 48){$\cdots$}

\put(290,40){\circle{4}}  
\put(270,48){$n-2$} 

\put(310,40){\circle{4}}  
\put(306,48){$n-1$} 

\put(292,40){\vector(1,0){16}}

\put(290,20){\circle{4}} 
\put(286,6){$n$} 

\put(290,38){\vector(0,-1){16}} 

\put(340,40){$= \ M(C)$}

\end{picture}
\end{center}
The trident graph $M(C)$ thus obtained is obviously invariant under a transposition $(n-1,n)$ of vertices. 
With this in mind, we define a {\it reflection} 
$R_{C}$ associated with a cycle graph $C$ 
by the following sequence of mutations and a transposition: 
\begin{equation} \label{eq:ref}
R_C=M^{-1} \circ (n-1,n) \circ M,
\quad M=\mu_{n-1, \ldots,2,1}=\mu_{n-1} \circ \cdots \circ \mu_2 \circ \mu_{1}.
\end{equation}
Then $R_C$ keeps $C$ invariant as
$R_C(C)= M^{-1}\circ (n-1,n) \circ M(C)=
 M^{-1}\circ  M(C)=C$.
It is immediate from ${\mu_k}^2=(n-1,n)^2={\rm id}$
that
 ${R_C}^2={\rm id}$.

Next we consider a quiver $Q$ obtained from the cycle $C$ by adding a {\it copy} $n'$ of the vertex $n$:
\begin{center}
\begin{picture}(150,70)

\put(0,50){$Q=$}
\put(40,50){\circle{4}}  
\put(36,58){$1$} 
\put(42,50){\vector(1,0){16}}
\put(60,50){\circle{4}}  
\put(56,58){$2$} 
\put(62,50){\vector(1,0){16}}
\put(80,50){\circle{4}}  
\put(76,58){$3$} 
\put(82,50){\vector(1,0){36}}
\put(92, 58){$\cdots$}
\put(120,50){\circle{4}}  
\put(116,58){$n-1$} 

\put(80,30){\circle{4}} 
\put(76,19){$n$} 
\put(78,31){\vector(-2,1){36}}  
\put(118,49){\vector(-2,-1){36}} 

\put(80,10){\circle{4}} 
\put(76,-4){$n'$} 
\put(78,11){\vector(-1,1){37}}  
\put(119,48){\vector(-1,-1){37}}  
\end{picture}
\end{center}
In this case,
applying the same composition $M=\mu_{n-1, \ldots,2,1}$ of mutations as above to $Q$ gives us
a four-pronged graph 
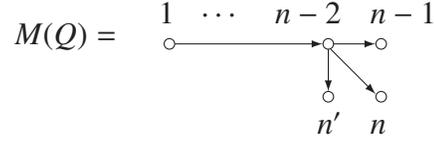
\begin{center}
\begin{picture}(150,50)

\put(-19,30){$M(Q)=$}
\put(40,30){\circle{4}}  
\put(36,38){$1$} 

\put(42,30){\vector(1,0){56}}

\put(52, 38){$\cdots$}

\put(100,30){\circle{4}}  
\put(80,38){$n-2$} 

\put(120,30){\circle{4}}  
\put(116,38){$n-1$} 

\put(102,30){\vector(1,0){16}}

\put(100,10){\circle{4}} 
\put(96,-4){$n'$} 

\put(100,28){\vector(0,-1){16}} 

\put(120,10){\circle{4}}
\put(101,28){\vector(1,-1){17}} 
\put(116,-4){$n$} 
\end{picture}
\end{center}
in which the three vertices  $n-1$, $n$  and $n'$  are symmetric.
Therefore $Q$ is invariant under the reflection 
$R_C$ defined by (\ref{eq:ref}).

\begin{prop}
 \label{prop:copy}
 $(R_C \circ (n,n'))^3={\rm id}.$ 
\end{prop}

\pf 
Since $M=\mu_{n-1,\ldots,2,1}$ and $(n,n')$ mutually commute,
it holds that
\begin{align*}
R_C\circ (n,n')&=M^{-1} \circ (n-1,n) \circ M \circ (n,n')
\\
&=M^{-1} \circ (n-1,n) \circ  (n,n') \circ M.
\end{align*} 
Noticing that $(n-1,n) \circ  (n,n')$ is a cyclic permutation of order three,
we arrive at the conclusion.
\qed

\subsection{Characterization of a quiver invariant under the reflections}
In general, 
when does a quiver $Q$ containing a cycle subgraph $C$ become invariant under the reflection $R_C$?
First we note that
$Q$ is invariant under $R_C$ if and only if $M(Q)$ is invariant under $(n,n-1)$.
Recall the definition (\ref{eq:ref}) of $R_C$.

The following lemma is elementary but crucial to solving this problem.

\begin{lemma}   \label{lemma:wedge}
Let  $Q=C \cup w$ be a quiver obtained from an $n$-cycle $C=(1 \to 2 \to \cdots \to n \to 1)$   by adding a `wedge graph'  $w=(k \to e \to \ell)$, 
where  $e$  is a new vertex 
and  $\{k, \ell\}$  are any two distinct vertices in $C$.  
Then  $R_C(Q)=Q$.
\end{lemma} 

\pf
The mutated quiver $M(Q)$ with $M=\mu_{n-1, \ldots,2,1}$
is invariant under $(n,n-1)$,
which can be easily verified by chasing the process of mutations.
See Example~\ref{ex:MQ} below. 
\qed

\begin{example} \label{ex:MQ}
\rm
If $2 \leq \ell <k \leq n-1$,  
the mutated quiver $M(Q)$ is obtained as follows:
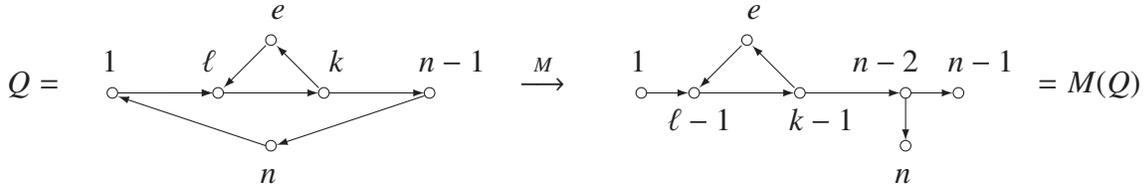
\begin{center}
\begin{picture}(430,70)

\put(0,30){$Q=$}
\put(40,30){\circle{4}}  
\put(36,38){$1$} 
\put(42,30){\vector(1,0){36}}

\put(62,30){\vector(1,0){16}}
\put(80,30){\circle{4}}  
\put(74,38){$\ell$}  
\put(82,30){\vector(1,0){36}}

\put(120,30){\circle{4}}  
\put(122,38){$k$}  

\put(156,38){$n-1$} 

\put(122,30){\vector(1,0){36}}
\put(160,30){\circle{4}}

\put(100,10){\circle{4}} 
\put(96,-4){$n$} 
\put(98,10){\vector(-3,1){56}}  
\put(158,29){\vector(-3,-1){56}}

\put(98,48){\vector(-1,-1){16}}
\put(118,32){\vector(-1,1){16}}

\put(100,50){\circle{4}} 
\put(100,58){$e$} 

\put(194,30){$\stackrel{M}{\longrightarrow} $}

\put(240,30){\circle{4}}  

\put(260,30){\circle{4}}  

\put(300,30){\circle{4}}  

\put(340,30){\circle{4}}  
\put(360,30){\circle{4}}  

\put(340,10){\circle{4}}  

\put(280,50){\circle{4}}  

\put(280,58){$e$} 

\put(278,48){\vector(-1,-1){16}}
\put(298,32){\vector(-1,1){16}}

\put(242,30){\vector(1,0){16}}
\put(262,30){\vector(1,0){36}}

\put(302,30){\vector(1,0){36}}
\put(342,30){\vector(1,0){16}}
  
 \put(340,28){\vector(0,-1){16}}

\put(236,38){$1$} 

\put(250,16){$\ell-1$} 

\put(296,16){$k-1$}  
\put(320,38){$n-2$}
\put(356,38){$n-1$} 

\put(336,-4){$n$} 

\put(390,30){$= M(Q)$}
\end{picture}
\end{center}
Therefore, $M(Q)$ is invariant under $(n,n-1)$.
\end{example}

Moreover,  a quiver obtained from a cycle $C$ by adding any number of wedge graphs still remains invariant under $R_C$. 

\begin{lemma} \label{lemma:multi-wedge}
Let  $Q^{(m)}=C \cup\bigcup_{i=1}^m w_i$  be a quiver obtained from an $n$-cycle $C$  by adding $m$ wedge graphs  $w_i=(k_i \to e_i \to \ell_i)$, 
where each $e_i$  is a new vertex,
$\{k_i, \ell_i\}$  are any two distinct vertices in  $C$, and
duplications among newly added vertices and edges are 
permitted. 
Then  $R_C(Q^{(m)})=Q^{(m)}$.
\end{lemma} 

\pf
Both 
 $M(C \cup w_i)$  and  $M(Q^{(m)})$ 
 contain
 $M(C)$ as a subgraph. 
 Two graphs $\bigcup_{i=1}^m (M(C \cup w_i)  -M(C))$ and $M(Q^{(m)})- M(C)$ coincide except for edges among the new vertices
 $\{ e_1,e_2,\ldots,e_m\}$.
 Here, for a graph $G$ and its subgraph $H$, the symbol $G-H$
 of {\it graph-difference} denotes the graph obtained from $G$ by removing all the edges of $H$.
By Lemma~\ref{lemma:wedge}, $M(C \cup w_i)  -M(C)$
is invariant under a transposition 
$(n-1,n)$, and so is $M(Q^{(m)})$.
\qed
\\

The following general fact comes easily from the mutation rule 
(\ref{eq:mut_B}) 
of quivers and will be used later in the proof of Theorem~\ref{thm:R-inv}.

\begin{lemma}  \label{lemma:subgraph}
Let $Q=(V,E)$ be a quiver.
For a given subset $V_0\subset V$, let $Q'$ denote the quiver obtained from $Q$ by removing all the edges among vertices of $V_1=V \setminus V_0$.
Then, for any sequence of mutations $M=\mu_{i_1,i_2,\ldots,i_\ell}$ $(i_1,i_2,\ldots,i_\ell \in V_0)$, two quivers
$M(Q)$ and $M(Q')$ coincide except for edges among vertices of $V_1$. 
\end{lemma}

A subgraph $H$ of a graph $G$ is called an {\it induced subgraph} 
if its edge set consists of all the edges of $G$ whose endpoints both belong to its vertex set.
We often write $H=G[U]$, where $U$ denotes the vertex set of $H$, 
because
an induced subgraph is uniquely determined by its vertex set.

Now we are ready to state a characterization of a quiver invariant under the reflections.

\begin{thm}
\label{thm:R-inv}
A quiver $Q$ which contains an $n$-cycle $C=(1 \to 2 \to \cdots \to n \to 1)$ as an induced subgraph is $R_C$-invariant if and only if 
the following condition holds{\rm:}
\begin{quote}
{\rm (W)}  \quad  \begin{tabular}{l}
For any vertex $v$ of $Q$ outside of $C$, the numbers of edges from $v$ to $C$  
\\ 
and  from $C$ to $v$ are equal. 
\end{tabular}
\end{quote}
\end{thm} 
\

In other words, the condition (W) means that any vertex of $Q$ outside of $C$ connects to the cycle subgraph $C$ with wedge graphs.
Note that the {\it sufficiency} of (W) has already been proved by Goncharov--Shen \cite[Section 7]{GS}.  \\

\pf
Let $Q'$ denote the subgraph of $Q$ consisting of all the edges connected to the vertices of $C$.
 
First we show the sufficiency.
 If $Q$ satisfies the condition (W) then $Q'$ 
 is a quiver obtained from $C$ by adding an appropriate number of wedge graphs.
By Lemma~\ref{lemma:multi-wedge},
 $M(Q')$ is  invariant under a transposition
$(n-1,n)$ of vertices, and so is $M(Q)$ via Lemma~\ref{lemma:subgraph}.
Hence $Q$ is $R_C$-invariant. 

Next we show the necessity. 
Assume,  for the sake of contradiction, 
 that an $R_C$-invariant quiver $Q$ does not satisfy (W).
If we remove (or add) wedge graphs appropriately from $Q'$
then we get a quiver $Q''$ 
that can be obtained from $C$
by adding some multiple edges $v_j \stackrel{m_j}{\to} n$ or $v_j \stackrel{m_j}{\leftarrow} n$ $(1 \leq j \leq p)$ to a {\it single} vertex $n$ of $C$,
where $v_1, v_2, \ldots, v_p$ are distinct vertices of $Q$ outside of $C$. 
Obviously, $M(Q'')$ is not symmetric with respect to $(n-1,n)$.
Because  $Q'$ is a quiver 
obtained from $Q''$ by adding an  appropriate number of wedge graphs  $w_1,w_2,\ldots,w_q$, the two graphs  $M(Q')-M(C)$ and $\bigcup_{k=1}^q (M(C \cup w_k)-M(C)) \cup (M(Q'')-M(C))$ coincide
except for the edges away from vertices of $C$. 
Consequently,
 $M(Q')$ is also not symmetric with respect to $(n-1,n)$.

 On the other hand, since $Q$ is $R_C$-invariant, $M(Q)$ is symmetric with respect to $(n-1,n)$; therefore, so is $M(Q')$ via Lemma~\ref{lemma:subgraph}.
 This is a contradiction.
\qed

\subsection{Explicit form of the birational transformation $R_C$ and its rotational symmetry}

For an $n$-cycle $C = (1\to 2\to \cdots\to n \to1)$,
we introduce the polynomials
\[
F_k=
F_k(y_1,y_2,\ldots, y_n)
=1+\sum_{j=1}^{n-1} \prod_{i=1}^{j} y_{k+i}
\quad (1 \leq k \leq n )
\]
 in $n$ variables,
regarding the suffix $i$ of the variable $y_i$ as an element of ${\mathbb Z}/n{\mathbb Z}$,  i.e. $y_{i+n}=y_i$.
For notational simplicity, we also extend the suffix $k$ of the polynomial for
any $k \in {\mathbb Z}$ by 
$F_{k+n}=F_k$.

We start with
a quiver $Q= C \cup w$
obtained from an $n$-cycle $C=(1 \to 2 \to \cdots \to n \to 1)$ by adding a wedge graph $w=(1 \to e \to n)$.
It follows from
the mutation rule (\ref{eq:mut_y})
that the action of 
$R_{C}=\mu_{1,2,\ldots,n-1}\circ (n-1,n) \circ \mu_{n-1,\ldots,2,1}$ on the variables $y_i$ $(1 \leq i \leq n)$ attached to the cycle subgraph $C$ 
coincides with the case where $Q=C$.
Note that the same is also true for any case where $Q$ is a quiver containing $C$ and is $R_C$-invariant. 

\begin{lemma}  \label{lemma:R_e} 
Let $Q=C \cup w$ be a quiver obtained from an $n$-cycle $C=(1 \to 2 \to \cdots \to n \to 1)$
by adding a wedge graph $w=(1 \to e \to n)$.
Then the birational action of $R_{C}$ on $y$-variables reads as 
\begin{equation} \label{eq:R_y}
R_C(y_i)=\frac{F_{i-1}}{y_{i+1} F_{i+1}}  \quad (1 \leq i  \leq n)
\end{equation}
and
\begin{equation} \label{eq:R_e}
R_C(y_e)= y_e\frac{y_1 F_1}{F_n}.
\end{equation}
\end{lemma} 

\pf
We prove it by  induction on the length $n$ of a cycle $C$.
When $n=2$, it is straightforward from (\ref{eq:mut_y}).

Suppose $n >2$.
A mutated quiver $Q'=\mu_1(Q)$ contains an $(n-1)$-cycle
$C'=( 2 \to 3 \to \cdots \to n \to 2)$
whose vertex set is $\{2,3, \ldots,n\}$:
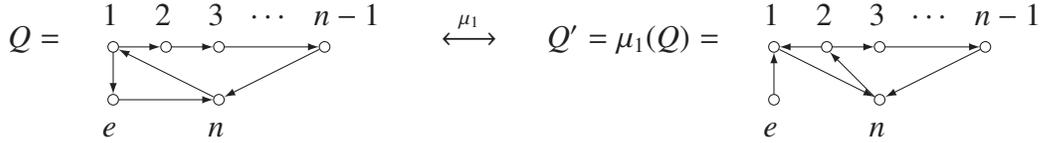
\begin{center}
\begin{picture}(400,50)

\put(0,30){$Q=$}
\put(40,30){\circle{4}}  
\put(36,38){$1$} 
\put(42,30){\vector(1,0){16}}
\put(60,30){\circle{4}}  
\put(56,38){$2$} 

\put(62,30){\vector(1,0){16}}
\put(80,30){\circle{4}}  
\put(76,38){$3$}  
\put(82,30){\vector(1,0){36}}
\put(92, 38){$\cdots$}
\put(120,30){\circle{4}}  
\put(116,38){$n-1$} 

\put(80,10){\circle{4}} 
\put(76,-4){$n$} 
\put(78,11){\vector(-2,1){36}}  
\put(118,29){\vector(-2,-1){36}} 

\put(40,10){\circle{4}} 
\put(40,28){\vector(0,-1){16}} 
\put(42,10){\vector(1,0){36}}
\put(36,-4){$e$} 

\put(164,30){$\stackrel{\mu_1}{\longleftrightarrow} $}

\put(204,30){$Q'= \mu_1(Q)=$}

\put(290,30){\circle{4}}  
\put(286,38){$1$} 
\put(308,30){\vector(-1,0){16}}
\put(310,30){\circle{4}}  
\put(306,38){$2$} 

\put(312,30){\vector(1,0){16}}
\put(330,30){\circle{4}}  
\put(326,38){$3$}  
\put(332,30){\vector(1,0){36}}
\put(342, 38){$\cdots$}
\put(370,30){\circle{4}}  
\put(366,38){$n-1$} 

\put(330,10){\circle{4}} 
\put(326,-4){$n$} 
\put(292,29){\vector(2,-1){36}}  

\put(368,29){\vector(-2,-1){36}} 

\put(290,10){\circle{4}} 
\put(290,12){\vector(0,1){16}}
\put(286,-4){$e$} 
\put(328,11){\vector(-1,1){17}} 

\end{picture}
\end{center}
It holds that 
$R_C=\mu_1 \circ R_{C'} \circ \mu_1$,
where $R_{C'}=\mu_{2,3,\ldots,n-1}\circ (n-1,n) \circ \mu_{n-1,\ldots,3,2}$ is the reflection associated with 
the $(n-1)$-cycle $C'$.
According to Theorem~\ref{thm:R-inv},  $Q'$ is $R_{C'}$-invariant.

Now, let us chase how the Y-seed
$\left(Q, {\boldsymbol y}=(y_1,y_2,\ldots,y_n,y_e)\right)$ is transformed at each step:
\[
(Q, {\boldsymbol y}) \stackrel{\mu_1}{\longleftrightarrow} 
(Q'=\mu_1(Q), {\boldsymbol y'}) \stackrel{R_{C'}}{\longleftrightarrow} 
(Q''=Q',{\boldsymbol y''})
\stackrel{\mu_1}{\longleftrightarrow}
(Q'''=Q,{\boldsymbol y'''}).
\]
By the induction hypothesis (see (\ref{eq:R_y}) and (\ref{eq:R_e})),
it holds that
\[
y''_1 =y'_1 \frac{y'_2F'_2}{F'_n},
\quad
y''_2=\frac{F'_{n}}{y'_{3} F'_{3}},
\quad
y''_k =\frac{F'_{k-1}}{y'_{k+1} F'_{k+1}}
\quad
(3 \leq k \leq n-1),
\quad
y''_n = \frac{F'_{n-1}}{y'_{2} F'_{2}},
\]
where we let
\[
F'_k =F'_k(y'_2,y'_3,\ldots, y'_n)
=1+\sum_{j=1}^{n-2} \prod_{i=1}^{j} y'_{k+i}
\quad (2 \leq k \leq n)
\]
regarding the suffix $i$ of the variable $y'_i$ as an element of ${\mathbb Z}/(n-1){\mathbb Z}$, i.e. $y'_{i+n-1}=y'_i$.
Moreover, it holds that $y_e''=R_{C'}(y_e')=y_e'$
since the vertex $e$ is not adjacent to the cycle $C'$ in the quiver $Q'=\mu_1(Q)$.
It follows from (\ref{eq:mut_y}) that
\[
y'_1= \frac{1}{y_1},\quad y'_2=y_2 \frac{y_1}{1+y_1}, \quad y'_k=y_k \quad (3 \leq k \leq n-1),
\quad y'_n =y_n(1+y_1),
\quad
y_e'=y_e \frac{y_1}{1+y_1}
\]
and, thereby, $F_k'=F_k$ $(2 \leq k \leq n-1)$ and $F'_n=F_n/(1+y_1)$.
Similarly, it holds that
\[
y'''_1=\frac{1}{y''_1},
\quad
y'''_2=y''_2(1+y''_1),
\quad y'''_k=y''_k \quad (3 \leq k \leq n-1),
\quad
y'''_n=y''_n \frac{y''_1}{1+y''_1},
\quad
y_e'''
=y''_e (1+y_1'').
\]
Combining these formulae
with the aid of an identity 
\[
F_k+y_{k+2} F_{k+2} = (1+y_{k+1})F_{k+1}  
\quad  (k \in {\mathbb Z}/n{\mathbb Z})
\]
leads to the desired results
 (\ref{eq:R_y}) and (\ref{eq:R_e}).
\qed
\\

If we apply $\mu_n$ to the same quiver $Q=C \cup w$ as above,
the resulting quiver $\check{Q}=\mu_n(Q)$
contains an $(n-1)$-cycle $\check{C}=(1 \to 2 \to \cdots \to n-1 \to 1)$
and the vertex $e$ is not adjacent to $\check C$ as follows:
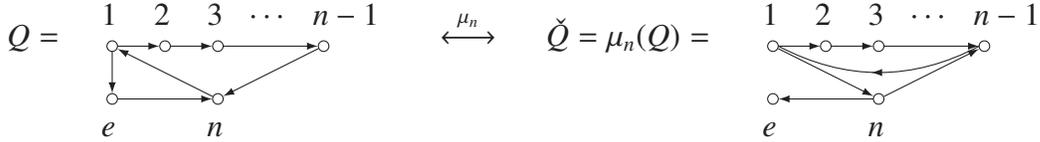
\begin{center}
\begin{picture}(400,50)

\put(0,30){$Q=$}
\put(40,30){\circle{4}}  
\put(36,38){$1$} 
\put(42,30){\vector(1,0){16}}
\put(60,30){\circle{4}}  
\put(56,38){$2$} 

\put(62,30){\vector(1,0){16}}
\put(80,30){\circle{4}}  
\put(76,38){$3$}  
\put(82,30){\vector(1,0){36}}
\put(92, 38){$\cdots$}
\put(120,30){\circle{4}}  
\put(116,38){$n-1$} 

\put(80,10){\circle{4}} 
\put(76,-4){$n$} 
\put(78,11){\vector(-2,1){36}}  
\put(118,29){\vector(-2,-1){36}} 

\put(40,10){\circle{4}} 
\put(40,28){\vector(0,-1){16}} 
\put(42,10){\vector(1,0){36}}
\put(36,-4){$e$} 

\put(164,30){$\stackrel{\mu_n}{\longleftrightarrow} $}

\put(204,30){$\check Q= \mu_n(Q)=$}

\put(290,30){\circle{4}}  
\put(286,38){$1$} 
\put(292,30){\vector(1,0){16}}
\put(310,30){\circle{4}}  
\put(306,38){$2$} 

\put(312,30){\vector(1,0){16}}
\put(330,30){\circle{4}}  
\put(326,38){$3$}  
\put(332,30){\vector(1,0){36}}
\put(342, 38){$\cdots$}
\put(370,30){\circle{4}}  
\put(366,38){$n-1$} 

\put(330,10){\circle{4}} 
\put(326,-4){$n$} 
\put(292,29){\vector(2,-1){36}}  

\put(332,11){\vector(2,1){36}} 

\put(290,10){\circle{4}} 
\put(328,10){\vector(-1,0){36}} 
\put(286,-4){$e$} 
\qbezier(292, 30)(330, 10)(368, 30)
\put(328,20){\vector(-1,0){1}} 

\end{picture}
\end{center}
Consider a sequence of mutations and a transposition
\[
T
= \mu_{n,1,2, \ldots, n-2} \circ (n-2,n-1) \circ \mu_{n-2, \ldots, 2,1,n} 
= \rho^{-1} \circ R_C \circ \rho
\]
with
$\rho =(1,2, \ldots,n) \in {\mathfrak S}_n$
being a cyclic permutation.
Then, as mentioned above, $T(y_i)=R_C(y_i)$ $(1 \leq i \leq n)$ holds
and, surprisingly, 
\[
T(y_e)=R_C(y_e)=y_e\frac{y_1 F_1}{F_n}
\]
also holds.
Therefore, the reflection $R_{C}$ still possesses  the rotational symmetry 
$\rho^{-1} \circ R_C \circ \rho=R_C$.

Furthermore, by repeatedly using this rotational symmetry, 
Lemma~\ref{lemma:R_e} is generalized to the next lemma
(the original case is $k=1$).

\begin{lemma}    \label{lemma:R_e2}  
Let $Q=C \cup w_k$ be a quiver obtained from an $n$-cycle $C=(1 \to 2 \to \cdots \to n \to 1)$
by adding a wedge graph $w_k=(k \to e \to k-1)$.
Then the birational action of the reflection $R_{C}$ on $y$-variables 
is given by 
{\rm(\ref{eq:R_y})} and
\[
R_C(y_e)= y_e\frac{y_{k} F_k}{F_{k-1}}.
\]
\end{lemma}

\begin{dfn} \rm  
A cycle subgraph $C$ of a quiver $Q=(V,E)$
is called a {\it balanced cycle} if
$C$ is an induced subgraph and satisfies
the condition (W) of Theorem~\ref{thm:R-inv}.
\end{dfn}

Hereafter,
when we consider a reflection $R_C$ associated with a cycle subgraph $C \subseteq Q$,
we assume that $C$ is a balanced cycle and thus $R_C(Q)=Q$.
For a balanced cycle $C=(1 \to 2 \to \cdots \to n \to 1)$
whose vertex set is $I=\{1,2,\ldots,n\}$
and a vertex $e$ outside of $C$, there exists
an $n$-tuple 
${\boldsymbol m}=(m_1,m_2,\ldots,m_n) \in {({\mathbb Z}_{\geq 0})}^n$
of nonnegative integers such that
 the induced subgraph $Q[\{e\} \cup I]$
 is identical to a quiver 
$C \cup \bigcup_{k=1}^n m_k w_k$ obtained from $C$
by adding a wedge graph $w_k=(k \to e \to k-1)$ with multiplicity $m_k$
for each $k \in I$.
We can and will normalize ${\boldsymbol m}$ by
$\min \{m_i \}_{i \in I}=0$.

\begin{prop}[cf. {\cite[Theorem 7.7]{GS}}]
\label{prop:birat}
Let $C=(1 \to 2 \to \cdots \to n \to 1)$ be a balanced $n$-cycle.
Then the birational action of the reflection $R_C$ on $y$-variables reads as follows{\rm:}
\\ 
\indent
{\rm (i)} for a vertex $i \in I=\{1,2,\ldots,n\}$ of $C$
\[
R_C(y_i)=\frac{F_{i-1}}{y_{i+1} F_{i+1}};
\] 
\indent
{\rm (ii)} for a vertex  $e$ adjacent to $C$ from outside
\begin{equation*}   \label{eq:R_e3}
R_C(y_e)= y_e \prod_{k=1}^n \left(\frac{y_{k} F_k}{F_{k-1}}\right)^{m_k},
\end{equation*}
\indent
where ${\boldsymbol m}=(m_1,m_2,\ldots,m_n) \in {({\mathbb Z}_{\geq 0})}^n$ is  chosen as above{\rm;}
\\  
\indent
{\rm (iii)} $R_C(y_v)=y_v$ for any other vertex $v$.  
\end{prop}

\pf
(i) has already been proved
(see Lemma~\ref{lemma:R_e} and the preceding paragraph)
and (iii) is obvious from the mutation rule (\ref{eq:mut_y}).

To prove (ii) we first
consider a quiver 
$C \cup \bigcup_{k=1}^n \bigcup_{j=1}^{m_k} w_{k,j}$ 
obtained from $C$ by adding $m_k$ wedge graphs $w_{k,j}=(k \to e_{k,j} \to k-1)$ $(1 \leq j \leq m_k)$ for each $k \in I$,
where newly added $|{\boldsymbol m}|=m_1 +m_2+\cdots+m_n$  vertices 
$\{e_{k,j}\}$ are distinct.
By virtue of  Lemma~\ref{lemma:R_e2}, we have
\begin{equation} \label{eq:R_ekj}
R_C(y_{e_{k,j}})= y_{e_{k,j}} \frac{y_{k} F_k}{F_{k-1}}.
\end{equation}
Gluing the vertices $\{e_{k,j}\}$ together to make a single vertex $e$,
we obtain a quiver identical to the induced subgraph $Q[\{e\} \cup I]$. 
The $y$-variable attached to the vertex $e$ is defined by
\begin{equation} \label{eq:ye}
y_e=\prod_{k=1}^n \prod_{j=1}^{m_k} y_{e_{k,j}}. 
\end{equation}
Noticing that  
the birational action of $R_{C}$ on $y_e$ is the same
for both quivers $Q$ and $Q[\{e\} \cup I]$,
we verify (ii) immediately by (\ref{eq:R_ekj}) and (\ref{eq:ye}). 
\qed
\\

In \cite{GS},
a proof of Proposition~\ref{prop:birat} is given 
by induction on the length of a cycle subgraph.
Our proof employs an idea of gluing vertices and requires almost no computation.
It is quite different from theirs, so we have written it above without omission.

From the explicit formulae for the birational
transformation $R_{C}$,
we observe again that the {\it rotational symmetry}
\begin{equation} \label{eq:rotation}
\rho^{-1} \circ R_C \circ \rho=R_C, \quad \rho=(1,2,\ldots,n) \in {\mathfrak S}_n
\end{equation}
holds.
Moreover, we have the following.

\begin{cor} \label{cor:sym}
For any permutation  $\sigma \in {\mathfrak S}_n$,
it holds that $\sigma^{-1} \circ R_C \circ \sigma =R_C$.
\end{cor} 

\pf
For any vertex $i_1 \in I=\{1,2,\ldots, n\}$ of the balanced $n$-cycle
$C=(1 \to 2 \to \cdots \to n \to 1)$, 
the mutated quiver $Q'=\mu_{i_1}(Q)$ 
contains a balanced $(n-1)$-cycle $C'$ whose 
 vertex set is $I \setminus \{i_1\}$.
 The rotational symmetry (\ref{eq:rotation}) implies that $R_C=\mu_{i_1} \circ R_{C'} \circ \mu_{i_1}$. 
Repeating the same argument as above, 
we find that $R_C=\mu_{i_1,i_2,\ldots,i_{n-1} } \circ (i_{n-1}, i_n)  \circ \mu_{i_{n-1}, \ldots, i_2,i_1}$
for any permutation 
\[
\sigma=
\begin{pmatrix} 1 & 2 &\cdots & n \\
i_1 & i_2 &\cdots & i_n \end{pmatrix} 
\in {\mathfrak S}_n. 
\]
\qed
\\

Although the  rotational symmetry (\ref{eq:rotation}) is a special case of Corollary~\ref{cor:sym},   it is rather essential as revealed by the proof above.

\begin{remark}  \label{remark:history}
\rm
In the context of integrable systems,
the reflection (\ref{eq:ref}) first appeared in
a study of geometric $R$-matrices 
and cluster algebras
by Inoue--Lam--Pylyavskyy \cite{ILP}.
In a recent study of higher-dimensional Teichm\"uller spaces
by Inoue--Ishibashi--Oya \cite{IIO},
birational representations of Coxeter groups 
were derived from cluster algebras corresponding to weighted quivers.
The first appearance of the reflection (\ref{eq:ref}) 
was perhaps 
in Bucher's work \cite{Bucher} on cluster algebras arising from 
surface triangulations and, as mentioned above, 
its properties were subsequently investigated in detail by Goncharov--Shen 
\cite[Section 7]{GS}. 
\end{remark}

\section{Relations of reflections}  \label{sect:rel}

Suppose a quiver $Q$ contains two or more balanced cycles;
thereby, $Q$ is invariant under the reflections associated with them.
What relations do the reflections satisfy?
In this section we prove the relations among reflections for some specific configurations of cycles. 
The rotational symmetry (\ref{eq:rotation}) of the reflection
plays a crucial role in the argument.

\subsection{Two intersecting cycles}
First we consider the following quiver:
\begin{center}
\begin{picture}(150,100)

\put(-10,50){$Q=$}

\put(60,50){\circle{4}} 
\put(50,40){$a$}

\put(80,50){\circle{4}} 
\put(70,55){$e$}

\put(100,50){\circle{4}} 
\put(102,55){$c$}

\put(80,70){\circle{4}} 
\put(85,70){$b$}

\put(80,30){\circle{4}} 
\put(70,20){$d$}

\thicklines
{\color{red}
\put(62,50){\vector(1,0){16}}
\put(80,52){\vector(0,1){16}}
}{\color{blue}
\put(98,50){\vector(-1,0){16}}
\put(80,48){\vector(0,-1){16}}
}{\color{red}\put(56,50.5){\vector(1,0){2}}
\put(58,72){\oval(44,44)[l]}
\put(58,72){\oval(44,44)[tr]}
\put(52,68){$\circlearrowleft$}
\put(20,78){$C_1$}
}{\color{blue}
\put(104,49.5){\vector(-1,0){2}}
\put(102,28){\oval(44,44)[r]}
\put(102,28){\oval(44,44)[bl]}
\put(96,24){$\circlearrowleft$}
\put(126,14){$C_2$}}

\thinlines

\put(78,32){\vector(-1,1){16}}

\put(82,68){\vector(1,-1){16}}

\end{picture}
\end{center}
such that 
two balanced cycles  $C_1 =(\cdots \to a \to e \to b \to \cdots)$ and
$C_2=(\cdots \to c \to e \to d \to \cdots)$
intersect at a vertex $e$,
where the lengths of these cycles may differ from each other.
The existence of two edges
$b \to c$ and $d \to a$ guarantees that $C_i$ $(i=1,2)$
are balanced and, thereby,
the invariance of $Q$ with respect to the associated reflections $R_{C_i}$;
see Theorem~\ref{thm:R-inv}.
Applying the mutation $\mu_e$ at the 
crossing vertex $e$
to $Q$ gives us the quiver
\begin{center}
\begin{picture}(150,100)

\put(-30,50){$\mu_e(Q)=$}

\put(60,50){\circle{4}} 
\put(50,40){$a$}

\put(80,50){\circle{4}} 
\put(85,55){$e$}

\put(100,50){\circle{4}} 
\put(102,55){$c$}
\put(80,70){\circle{4}} 
\put(85,70){$b$}
\put(80,30){\circle{4}} 
\put(70,20){$d$}

\thicklines
{\color{red}
\put(62,52){\vector(1,1){16}}
}{\color{blue}
\put(98,48){\vector(-1,-1){16}}
}{\color{red}
\put(56,50.5){\vector(1,0){2}}
\put(58,72){\oval(44,44)[l]}
\put(58,72){\oval(44,44)[tr]}

\put(52,68){$\circlearrowleft$}
\put(20,78){$C'_1$}
}{\color{blue}
\put(104,49.5){\vector(-1,0){2}}

\put(102,28){\oval(44,44)[r]}
\put(102,28){\oval(44,44)[bl]}

\put(96,24){$\circlearrowleft$}
\put(126,14){$C'_2$}
}
\thinlines
\put(78,50){\vector(-1,0){16}}
\put(82,50){\vector(1,0){16}}
\put(80,68){\vector(0,-1){16}}
\put(80,32){\vector(0,1){16}}

\end{picture}
\end{center}
in which
two balanced cycles 
$C'_1 =(\cdots \to a \to  b \to \cdots)$ and  $C'_2=(\cdots \to c \to  d \to \cdots)$
are not adjacent.
Hence the reflections $R_{C'_1}$ and $R_{C'_2}$ mutually commute.
By virtue of 
the rotational symmetry (\ref{eq:rotation}), 
it holds that 
\begin{equation} \label{eq:commute_1}
R_{C_i}=\mu_e \circ R_{C'_i}\circ \mu_e \quad (i=1,2).
\end{equation}
Therefore, the commutativity of $R_{C_1}$ and $R_{C_2}$ is concluded as follows:
\begin{equation} \label{eq:commute_2}
R_{C_1} \circ R_{C_2}=\mu_e \circ R_{C'_1}\circ R_{C'_2} \circ \mu_e
=\mu_e \circ R_{C'_2}\circ R_{C'_1} \circ \mu_e
=R_{C_2} \circ R_{C_1}.
\end{equation}

\def\ul{\underline}

In general, 
no matter how many crossing vertices there are, 
the commutativity of the reflections associated with two intersecting cycles
is valid
by the same mechanism.
Let us consider a quiver $Q$ containing
two balanced cycles 
 $C=(1\to2\to\dots\to n\to1)$ and $\ul{C}=(\ul{1}\to\ul{2}\to\dots\to\ul{m}\to\ul{1})$ 
 whose vertex sets are  
$I=\{1,2,\ldots,n\}$
and 
$\underline{I}=\{  \underline{1},  \underline{2}, \ldots,  \underline{m}\}$, respectively.
 Suppose
$C$ and $\ul C$ intersect at $\ell$ vertices $u_i=\ul{v_i} \in I \cap \ul{I}$ $(1 \leq i \leq \ell)$;
two edges $u_i+1\to\ul{v_i-1}$ and $\ul{v_i+1}\to u_i-1$ are equipped for each crossing vertex $u_i=\ul{v_i}$
while there is no other edge between $C$ and $\ul{C}$.
We also impose the assumption that 
no two crossing vertices $u_i=\ul{v_i}$ and $u_j=\ul{v_j}$ $(i\ne j)$ 
 are adjacent, 
 i.e.  there are one or more vertices 
 between the crossing vertices on each $C$ and $\ul{C}$.
 The figure below illustrates the induced subgraph 
 $Q[I \cup \ul{I}] \subseteq Q$. 
\begin{center}
\begin{picture}(346,118)

\put(30,40){\circle{4}} 
\put(86,40){\circle{4}}
   \put(15,25){$\ul{v_1+1}$}\put(70,25){$\ul{v_1-1}$}
\put(140,40){\circle{4}} 
\put(196,40){\circle{4}} 
   \put(125,25){$\ul{v_2+1}$}\put(180,25){$\ul{v_2-1}$}  
\put(260,40){\circle{4}} 
\put(316,40){\circle{4}} 
   \put(245,25){$\ul{v_\ell+1}$}\put(300,25){$\ul{v_\ell-1}$}   
\put(58,54){\circle{4}} 
   \put(55,40){$\ul{v_1}$} \put(55,65){$u_1$}
\put(168,54){\circle{4}} 
   \put(165,40){$\ul{v_2}$} \put(165,65){$u_2$}
\put(288,54){\circle{4}} 
   \put(285,40){$\ul{v_\ell}$} \put(285,65){$u_\ell$} 
\put(30,68){\circle{4}} 
\put(86,68){\circle{4}} 
   \put(15,78){$u_1-1$}\put(70,78){$u_1+1$}
\put(140,68){\circle{4}} 
\put(196,68){\circle{4}} 
   \put(125,78){$u_2-1$}\put(180,78){$u_2+1$}
\put(260,68){\circle{4}} 
\put(316,68){\circle{4}} 
 \put(245,78){$u_\ell-1$}\put(300,78){$u_\ell+1$}

{\color{blue}{\thicklines
\put(56,53){\vector(-2,-1){24}}
\put(84,41){\vector(-2,1){24}}
\put(166,53){\vector(-2,-1){24}}
\put(194,41){\vector(-2,1){24}}
\put(286,53){\vector(-2,-1){24}}
\put(314,41){\vector(-2,1){24}}
}}{\color{red}{\thicklines
\put(32,67){\vector(2,-1){24}}
\put(60,55){\vector(2,1){24}}
\put(142,67){\vector(2,-1){24}}
\put(170,55){\vector(2,1){24}}
\put(262,67){\vector(2,-1){24}}
\put(290,55){\vector(2,1){24}}
}}{\color{red} 
{\thicklines
\put(88,68){\vector(1,0){50}}
\put(198,68){\vector(1,0){18}}
\put(221,65){$\cdots$}
\put(240,68){\vector(1,0){18}}
\put(28,83){\oval(45,30)[l]}
\put(318,83){\oval(45,30)[r]}
\put(28,98){\line(1,0){290}}
\put(168,98){\vector(-1,0){2}}
\put(26,68){\vector(1,0){2}}
}}{\color{blue} 
{\thicklines
\put(138,40){\vector(-1,0){50}}
\put(216,40){\vector(-1,0){18}}
\put(221,37){$\cdots$}
\put(258,40){\vector(-1,0){18}}
\put(28,25){\oval(45,30)[l]}
\put(318,25){\oval(45,30)[r]}
\put(28,10){\line(1,0){290}}
\put(173,10){\vector(1,0){2}}
\put(320,40){\vector(-1,0){2}}
}}
\put(30,42){\vector(0,1){24}}
\put(86,66){\vector(0,-1){24}}
\put(140,42){\vector(0,1){24}}
\put(196,66){\vector(0,-1){24}}
\put(260,42){\vector(0,1){24}}
\put(316,66){\vector(0,-1){24}}
{\color{red}
\put(170,107){$C$}}{\color{blue}
\put(170,-6){$\underline C$}}
 \end{picture}
\end{center}
 \ 
\begin{thm}  
\label{thm:commute} 
The reflections $R_C$ and $R_{\ul C}$ mutually commute.
\end{thm}

\pf
Applying the mutation at every crossing vertex $u_i=\ul{v_i}$ 
to separate two cycles $C$ and $\ul C$,
we can easily verify the statement via the rotational symmetry
as above; cf. (\ref{eq:commute_1}) and  (\ref{eq:commute_2}).
\qed

\begin{remark}\rm
Theorem 4.1 can be extended to a more general setting where crossing vertices may be placed consecutively.
To be precise,
the commutativity of $R_C$ and $R_{\ul C}$ still holds
if the neighborhood of each $i$th crossing vertex
$u_i$ $(=\ul{v_i})$: 
\begin{center}
\begin{picture}(120,80)
\put(40,20){\circle{4}} \put(21,3){$u_i-1$}  
\put(80,20){\circle{4}} \put(67,3){$\ul{v_i+1}$}
\put(78,20){\vector(-1,0){36}}

\put(60,40){\circle{4}}
\put(70,36){$u_i$}

\put(40,60){\circle{4}} \put(21,72){$\ul{v_i-1}$}  
\put(80,60){\circle{4}} \put(67,72){$u_i+1$}
\put(78,60){\vector(-1,0){36}}

{\color{red}{\thicklines
\put(110,56){$C$}
\put(2,16){$C$}
\put(18,20){\vector(1,0){20}}
\put(82,60){\vector(1,0){20}}
\put(42,22){\vector(1,1){16}}
\put(62,42){\vector(1,1){16}}
}}{\color{blue}{\thicklines
\put(110,16){$\underline{C}$}
\put(2,56){$\underline{C}$}
\put(18,60){\vector(1,0){20}}
\put(82,20){\vector(1,0){20}}
\put(42,58){\vector(1,-1){16}}
\put(62,38){\vector(1,-1){16}}
}}
\end{picture}
\end{center}
is replaced by the following:
\begin{center}
\begin{picture}(140,80)
\put(40,20){\circle{4}} \put(21,3){$u_{i,1}-1$}  
\put(100,20){\circle{4}} \put(80,3){$\ul{v_{i,k_i}+1}$}
\put(98,20){\vector(-1,0){56}}

\put(40,40){\circle{4}}  \put(21,36){$u_{i,1}$}
\put(60,40){\circle{4}} \put(53,27){$u_{i,2}$} 
                                  \put(72, 27){$\cdots$}
\put(100,40){\circle{4}} \put(108,36){$u_{i,k_i}$}

\put(40,60){\circle{4}} \put(21,72){$\ul{v_{i,1}-1}$}  
\put(100,60){\circle{4}} \put(80,72){$u_{i,k_i}+1$}
\put(98,60){\vector(-1,0){56}}

{\color{violet}{\thicklines
\put(42,40){\vector(1,0){16}}
\put(62,40){\vector(1,0){36}}
}}{\color{red}{\thicklines
\put(130,56){$C$}
\put(2,16){$C$}
\put(18,20){\vector(1,0){20}}
\put(102,60){\vector(1,0){20}}
\put(40,22){\vector(0,1){16}}
\put(100,42){\vector(0,1){16}}
}}{\color{blue}{\thicklines
\put(130,16){$\underline{C}$}
\put(2,56){$\underline{C}$}
\put(18,60){\vector(1,0){20}}
\put(102,20){\vector(1,0){20}}
\put(40,58){\vector(0,-1){16}}
\put(100,38){\vector(0,-1){16}}
}}
\end{picture}
\end{center}
where
$ \{ u_{i,j}=\underline{v_{i,j}} \  (j=1,2,\ldots,k_i)\}\subset I \cap \underline{I}$
is the segment of $k_i$ consecutive vertices
on $C$ and $\ul C$
with $k_i$ being an arbitrary positive integer.
It can be proved 
as well as 
Theorem~\ref{thm:commute}
by sequentially applying the mutations at the crossing vertices 
$I \cap \underline{I}$.
\end{remark}

\subsection{Two cycles connected with a hinge}
\label{subsec:hinge}

Let us consider a quiver $Q$
containing two balanced cycles
$C=(1 \to 2 \to \cdots \to n \to 1)$ and 
$\underline{C}=(\underline 1 \to \underline 2 \to \cdots \to \underline m \to  \underline 1 )$
whose vertex sets are  
$I=\{1,2,\ldots,n\}$
and 
$\underline{I}=\{  \underline{1},  \underline{2}, \ldots,  \underline{m}\}$, respectively.
Suppose $C$ and $\underline{C}$ are connected
with 
a `hinge' $(1 \to \underline{m} \to n \to \underline{1} \to 1)$
while there is no other edge between $I$ and $\underline I$.

\begin{thm} \label{thm:hinge}
$(R_C \circ R_{\underline{C}})^3= {\rm id}$.
\end{thm} 

\pf
It is sufficient to be concerned with the induced subgraph $Q[I \cup \ul{I}] \subseteq Q$. 

If $n=2$, i.e. $C=(1 \to 2 \to 1)$, then the mutated quiver $\mu_1(Q[I \cup \ul{I}])$ 
takes the form of
an $(m+1)$-cycle $\underline{C}'=(1 \to \underline 1 \to \underline 2 \to \cdots \to \underline m \to  1 )$ 
with 
a copy $2$ of the vertex  $1$ added:
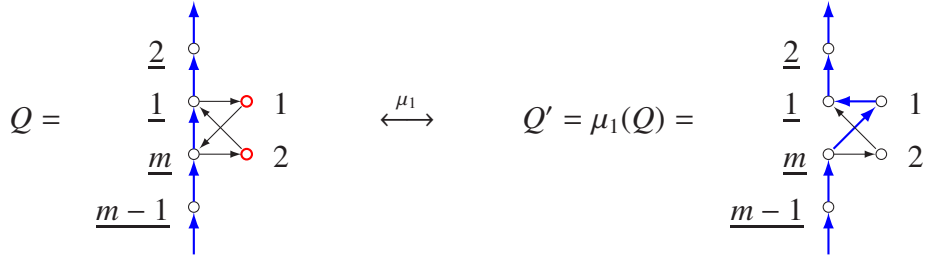
\begin{center}
\begin{picture}(350,100)

\put(-15,50){$Q[I \cup \ul{I}] =$}

\put(80,20){\circle{4}}
\put(80,40){\circle{4}}
\put(80,60){\circle{4}}
\put(80,80){\circle{4}}

{\thicklines\color{red}
\put(100,60){\circle{4}}
\put(100,40){\circle{4}}
}

\put(63,75){$\underline{2}$}
\put(63,55){$\underline{1}$}
\put(63,35){$\underline{m}$}
\put(43,15){$\underline{m-1}$}

\put(110,55){$1$}
\put(110,35){$2$}

{\thicklines\color{blue}
\put(80,2){\vector(0,1){16}}
\put(80,22){\vector(0,1){16}}
\put(80,42){\vector(0,1){16}}
\put(80,62){\vector(0,1){16}}
\put(80,82){\vector(0,1){16}}
}


\put(82,60){\vector(1,0){16}}
\put(82,40){\vector(1,0){16}}


\put(98,58){\vector(-1,-1){16}}
\put(98,42){\vector(-1,1){16}}

\put(150,50){$\stackrel{\mu_1}{\longleftrightarrow} $}

\put(204,50){$\mu_1(Q[I \cup \ul{I}])=$}

\put(320,20){\circle{4}}
\put(320,40){\circle{4}}
\put(320,60){\circle{4}}
\put(320,80){\circle{4}}

\put(340,40){\circle{4}}
\put(340,60){\circle{4}}

\put(303,75){$\underline{2}$}
\put(303,55){$\underline{1}$}
\put(303,35){$\underline{m}$}
\put(283,15){$\underline{m-1}$}

\put(350,55){$1$}
\put(350,35){$2$}

{\thicklines\color{blue}
\put(320,2){\vector(0,1){16}}
\put(320,22){\vector(0,1){16}}
\put(320,62){\vector(0,1){16}}
\put(320,82){\vector(0,1){16}}
\put(338,60){\vector(-1,0){16}}
\put(322,42){\vector(1,1){16}}
}

\put(322,40){\vector(1,0){16}}

\put(338,42){\vector(-1,1){16}}

\end{picture}
\end{center}
Therefore, Proposition~\ref{prop:copy} tells us 
that $(R_{\underline{C}'} \circ (1,2))^3={\rm id}$.
By virtue of the rotational symmetry (\ref{eq:rotation}), it holds that
$R_{\underline C}= \mu_1 \circ R_{\underline{C}'} \circ \mu_1$. 
Combining this with the definition $R_C=\mu_1 \circ (1,2)\circ \mu_1$ of the reflection,
we conclude that $(R_{C} \circ R_{\underline{C}})^3={\rm id}$.

If $n >2$ then
the mutated quiver $\mu_1(Q[I \cup \ul{I}])$ 
takes the form of two balanced cycles
$C'=(2 \to 3 \to \cdots \to n \to 2)$ and
$\underline{C}'=(1 \to \underline 1 \to \underline 2 \to \cdots \to \underline m \to  1 )$ 
of length $n-1$ and $m+1$, respectively, which are connected with a hinge
$(2 \to 1 \to n \to \underline{1} \to 2)$:
\begin{center}
\begin{picture}(440,100)

\put(-15,50){$Q[I \cup \ul{I}]=$}

\put(70,20){\circle{4}}
\put(70,40){\circle{4}}
\put(70,60){\circle{4}}
\put(70,80){\circle{4}}

\put(90,80){\circle{4}}
\put(90,60){\circle{4}}
\put(90,40){\circle{4}}
\put(90,20){\circle{4}}

\put(53,75){$\underline{2}$}
\put(53,55){$\underline{1}$}
\put(53,35){$\underline{m}$}
\put(33,15){$\underline{m-1}$}

\put(100,75){$2$}
\put(100,55){$1$}
\put(100,35){$n$}
\put(100,15){$n-1$}

{\thicklines\color{blue}
\put(70,2){\vector(0,1){16}}
\put(70,22){\vector(0,1){16}}
\put(70,42){\vector(0,1){16}}
\put(70,62){\vector(0,1){16}}
\put(70,82){\vector(0,1){16}}
}{\thicklines\color{red}
\put(90,2){\vector(0,1){16}}
\put(90,22){\vector(0,1){16}}
\put(90,42){\vector(0,1){16}}
\put(90,62){\vector(0,1){16}}
\put(90,82){\vector(0,1){16}}
}


\put(72,60){\vector(1,0){16}}
\put(72,40){\vector(1,0){16}}


\put(88,58){\vector(-1,-1){16}}
\put(88,42){\vector(-1,1){16}}

\put(140,50){$\stackrel{\mu_1}{\longleftrightarrow} $}

\put(184,50){$\mu_1(Q[I \cup \ul{I}])=$}

\put(300,20){\circle{4}}
\put(300,40){\circle{4}}
\put(300,60){\circle{4}}
\put(300,80){\circle{4}}

\put(320,80){\circle{4}}
\put(320,60){\circle{4}}
\put(320,40){\circle{4}}
\put(320,20){\circle{4}}

\put(283,75){$\underline{2}$}
\put(283,55){$\underline{1}$}
\put(283,35){$\underline{m}$}
\put(263,15){$\underline{m-1}$}

\put(330,75){$2$}
\put(335,55){$1$}
\put(330,35){$n$}
\put(330,15){$n-1$}

{\thicklines\color{blue}
\put(300,2){\vector(0,1){16}}
\put(300,22){\vector(0,1){16}}
\put(300,62){\vector(0,1){16}}
\put(300,82){\vector(0,1){16}}
}{\thicklines\color{red}
\put(320,82){\vector(0,1){16}}
}
\put(320,78){\vector(0,-1){16}}
\put(320,58){\vector(0,-1){16}}
{\thicklines\color{red}
\put(320,22){\vector(0,1){16}}
\put(320,2){\vector(0,1){16}}
}{\thicklines\color{blue}
\put(318,60){\vector(-1,0){16}}
\put(302,42){\vector(1,1){16}}
}


\put(302,62){\vector(1,1){16}}

\put(318,42){\vector(-1,1){16}}


{\thicklines\color{red}
\qbezier(322,42)(340,60)(322,78)
\put(324,76){\vector(-1,1){2}}
}

\put(365,50){$=$}

\put(410,20){\circle{4}}
\put(410,40){\circle{4}}
\put(410,60){\circle{4}}
\put(410,80){\circle{4}}

\put(430,80){\circle{4}}
\put(430,60){\circle{4}}
\put(430,40){\circle{4}}
\put(430,20){\circle{4}}

\put(393,75){$\underline{2}$}
\put(393,55){$\underline{1}$}
\put(393,35){$1$}
\put(393,15){$\underline{m}$}

\put(440,75){$3$}
\put(440,55){$2$}
\put(440,35){$n$}
\put(440,15){$n-1$}

{\thicklines\color{blue}
\put(410,2){\vector(0,1){16}}
\put(410,22){\vector(0,1){16}}
\put(410,42){\vector(0,1){16}}
\put(410,62){\vector(0,1){16}}
\put(410,82){\vector(0,1){16}}
\color{red}
\put(430,2){\vector(0,1){16}}
\put(430,22){\vector(0,1){16}}
\put(430,42){\vector(0,1){16}}
\put(430,62){\vector(0,1){16}}
\put(430,82){\vector(0,1){16}}
}


\put(412,60){\vector(1,0){16}}
\put(412,40){\vector(1,0){16}}


\put(428,58){\vector(-1,-1){16}}
\put(428,42){\vector(-1,1){16}}

\end{picture}
\end{center}
By repeating the same procedure, i.e. by sequentially applying  the mutations $\mu_2, \mu_3, \ldots, \mu_{n-2}$ to $\mu_1(Q[I \cup \ul{I}])$,
it reduces to the case where $n=2$.
\qed

\subsection{Two adjacent cycles in a ladder shape}
\label{subsec:ladder}

We consider a quiver $Q$ containing two balanced cycles
$C=(1 \to 2 \to \cdots \to n \to 1)$ and 
$\underline{C}=(\underline 1 \to \underline 2 \to \cdots \to \underline n \to  \underline 1 )$
of the same length
whose vertex sets are  
$I=\{1,2,\ldots,n\}$
and 
$\underline{I}=\{  \underline{1},  \underline{2}, \ldots,  \underline{n}\}$, respectively.
Suppose $C$ and $\underline{C}$ are connected with
$n$ consecutive wedge graphs $\underline i \to i \to \underline{i-1}$ $(i \in {\mathbb Z}/n{\mathbb Z})$
like a ladder
while there is no other edge between $I$ and $\underline I$.

\begin{thm}
\label{thm:ladder}
$(R_C \circ R_{\underline{C}})^3= {\rm id}$.
\end{thm} 

\pf 
If $n=2$ then it reduces to the case of Theorem~\ref{thm:hinge}
with $m=n=2$.

Suppose $n >2$.
Let $Q^{(0)}$ denote the induced subgraph $Q[I \cup \ul{I}] \subseteq Q$. 
Apply the composition of mutations
$M=\mu_{\underline{n-2}, n-2, \ldots, \underline 2,2,\underline 1,1}$
to $Q^{(0)}$.
Then we can chase the mutated quivers as follows:
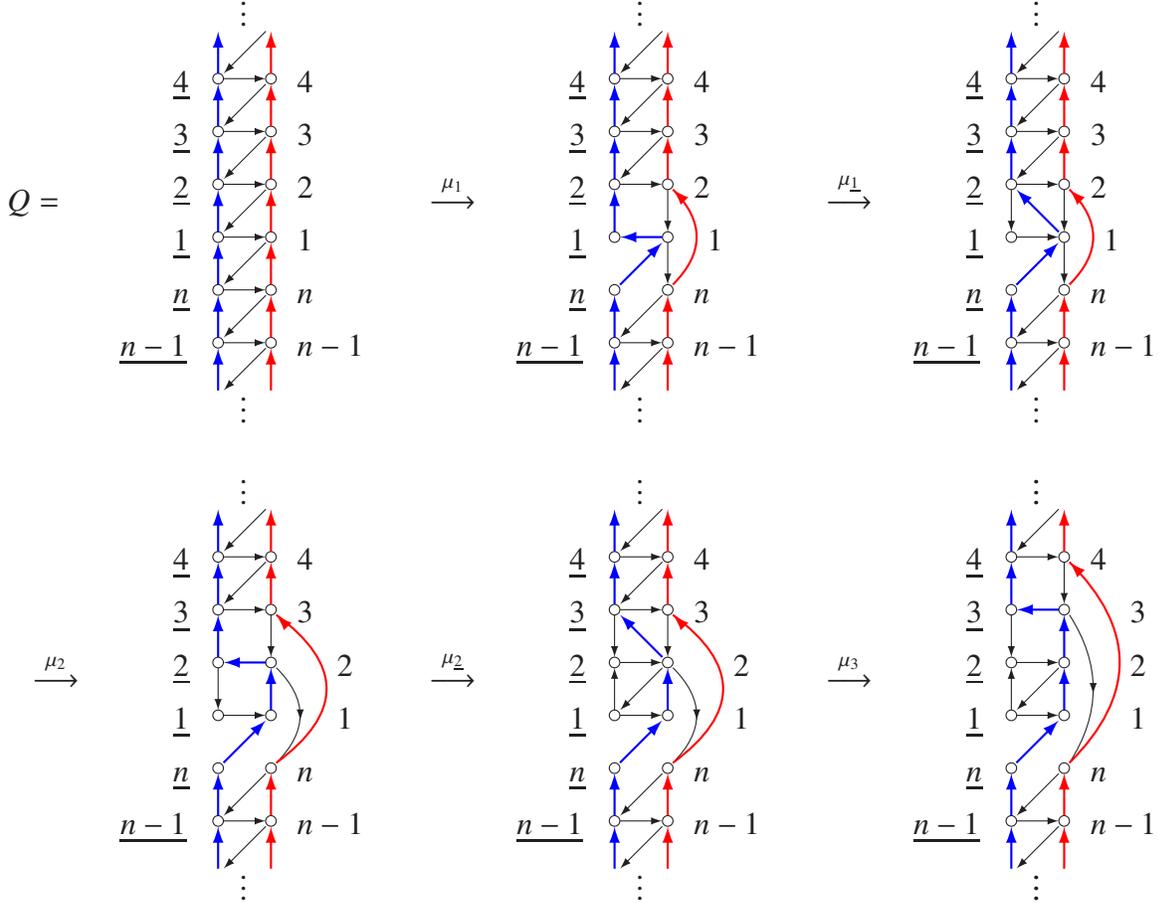
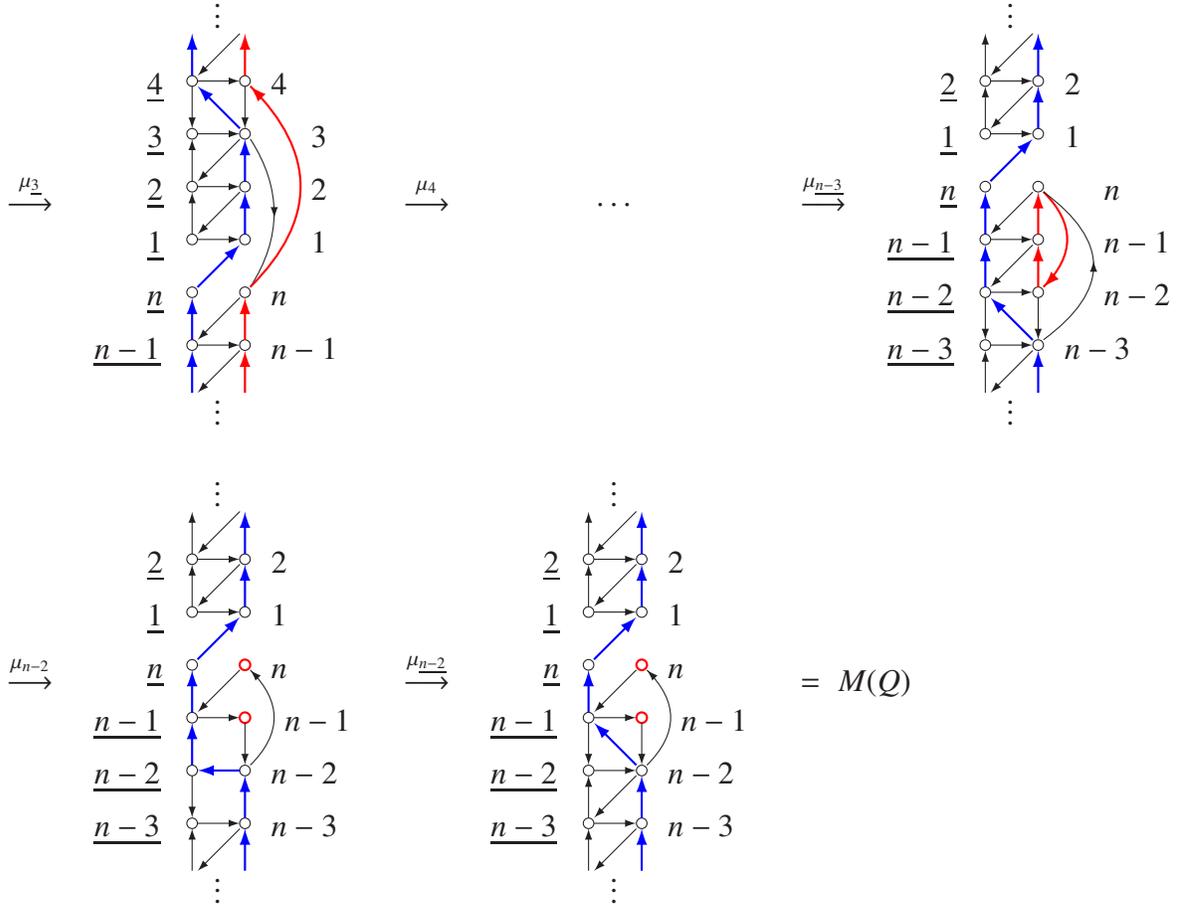
\begin{center}
\begin{picture}(430,180)

\put(0,90){$Q^{(0)}=$}

\put(80,40){\circle{4}}
\put(80,60){\circle{4}}
\put(80,80){\circle{4}}
\put(80,100){\circle{4}}
\put(80,120){\circle{4}}
\put(80,140){\circle{4}} 

\put(100,40){\circle{4}}
\put(100,60){\circle{4}}
\put(100,80){\circle{4}}
\put(100,100){\circle{4}}
\put(100,120){\circle{4}}
\put(100,140){\circle{4}} 

\put(63,135){$\underline{4}$}
\put(63,115){$\underline{3}$}
\put(63,95){$\underline{2}$}
\put(63,75){$\underline{1}$}
\put(63,55){$\underline{n}$}
\put(43,35){$\underline{n-1}$}

\put(110,135){$4$}
\put(110,115){$3$}
\put(110,95){$2$}
\put(110,75){$1$}
\put(110,55){$n$}
\put(110,35){$n-1$}
{\thicklines\color{blue}
\put(80,22){\vector(0,1){16}}
\put(80,42){\vector(0,1){16}}
\put(80,62){\vector(0,1){16}}
\put(80,82){\vector(0,1){16}}
\put(80,102){\vector(0,1){16}}
\put(80,122){\vector(0,1){16}}
\put(80,142){\vector(0,1){16}}
\color{red}
\put(100,22){\vector(0,1){16}}
\put(100,42){\vector(0,1){16}}
\put(100,62){\vector(0,1){16}}
\put(100,82){\vector(0,1){16}}
\put(100,102){\vector(0,1){16}}
\put(100,122){\vector(0,1){16}}
\put(100,142){\vector(0,1){16}}}

\put(82,140){\vector(1,0){16}}
\put(82,120){\vector(1,0){16}}
\put(82,100){\vector(1,0){16}}
\put(82,80){\vector(1,0){16}}
\put(82,60){\vector(1,0){16}}
\put(82,40){\vector(1,0){16}}

\put(98,158){\vector(-1,-1){16}}
\put(98,138){\vector(-1,-1){16}}
\put(98,118){\vector(-1,-1){16}}
\put(98,98){\vector(-1,-1){16}}
\put(98,78){\vector(-1,-1){16}}
\put(98,58){\vector(-1,-1){16}}
\put(98,38){\vector(-1,-1){16}}

\put(88,160){$\vdots$}
\put(88,10){$\vdots$}

\put(160,90){$\stackrel{\mu_1}{\longrightarrow} $}

\put(230,140){\circle{4}} 
\put(230,120){\circle{4}}
\put(230,100){\circle{4}}
\put(230,80){\circle{4}}
\put(230,60){\circle{4}}
\put(230,40){\circle{4}}

\put(250,140){\circle{4}} 
\put(250,120){\circle{4}}
\put(250,100){\circle{4}}
\put(250,80){\circle{4}}
\put(250,60){\circle{4}}
\put(250,40){\circle{4}}

\put(213,135){$\underline{4}$}
\put(213,115){$\underline{3}$}
\put(213,95){$\underline{2}$}
\put(213,75){$\underline{1}$}
\put(213,55){$\underline{n}$}
\put(193,35){$\underline{n-1}$}

\put(260,135){$4$}
\put(260,115){$3$}
\put(260,95){$2$}
\put(265,75){$1$}
\put(260,55){$n$}
\put(260,35){$n-1$}

{\thicklines\color{blue}
\put(230,142){\vector(0,1){16}}
\put(230,122){\vector(0,1){16}}
\put(230,102){\vector(0,1){16}}
\put(230,82){\vector(0,1){16}}
\put(230,42){\vector(0,1){16}}
\put(230,22){\vector(0,1){16}}
}{\thicklines\color{red}
\put(250,142){\vector(0,1){16}}
\put(250,122){\vector(0,1){16}}
\put(250,102){\vector(0,1){16}}
}{\thicklines\color{red}
\put(250,42){\vector(0,1){16}}
\put(250,22){\vector(0,1){16}}
}

\put(250,98){\vector(0,-1){16}}
\put(250,78){\vector(0,-1){16}}

\put(232,140){\vector(1,0){16}}
\put(232,120){\vector(1,0){16}}
\put(232,100){\vector(1,0){16}}
\put(232,40){\vector(1,0){16}}

{\thicklines\color{blue}
\put(248,80){\vector(-1,0){16}}
}

\put(248,158){\vector(-1,-1){16}}
\put(248,138){\vector(-1,-1){16}}
\put(248,118){\vector(-1,-1){16}}
\put(248,58){\vector(-1,-1){16}}
\put(248,38){\vector(-1,-1){16}}

{\thicklines\color{blue}
\put(232,62){\vector(1,1){16}}
}
\put(238,160){$\vdots$}
\put(238,10){$\vdots$}


{\thicklines\color{red}
\qbezier(252,62)(270,80)(252,98)
\put(254,96){\vector(-1,1){2}}
}

\put(310,90){$\stackrel{  \mu_{ \underline 1} }{\longrightarrow} $}


\put(380,140){\circle{4}} 
\put(380,120){\circle{4}}
\put(380,100){\circle{4}}
\put(380,80){\circle{4}}
\put(380,60){\circle{4}}
\put(380,40){\circle{4}}

\put(400,140){\circle{4}} 
\put(400,120){\circle{4}}
\put(400,100){\circle{4}}
\put(400,80){\circle{4}}
\put(400,60){\circle{4}}
\put(400,40){\circle{4}}

\put(363,135){$\underline{4}$}
\put(363,115){$\underline{3}$}
\put(363,95){$\underline{2}$}
\put(363,75){$\underline{1}$}
\put(363,55){$\underline{n}$}
\put(343,35){$\underline{n-1}$}

\put(410,135){$4$}
\put(410,115){$3$}
\put(410,95){$2$}
\put(415,75){$1$}
\put(410,55){$n$}
\put(410,35){$n-1$}

{\thicklines\color{blue}
\put(380,142){\vector(0,1){16}}
\put(380,122){\vector(0,1){16}}
\put(380,102){\vector(0,1){16}}
\put(380,42){\vector(0,1){16}}
\put(380,22){\vector(0,1){16}}
}

\put(380,98){\vector(0,-1){16}}

{\thicklines\color{red}
\put(400,142){\vector(0,1){16}}
\put(400,122){\vector(0,1){16}}
\put(400,102){\vector(0,1){16}}
\put(400,42){\vector(0,1){16}}
\put(400,22){\vector(0,1){16}}
}

\put(400,98){\vector(0,-1){16}}
\put(400,78){\vector(0,-1){16}}

\put(382,140){\vector(1,0){16}}
\put(382,120){\vector(1,0){16}}
\put(382,100){\vector(1,0){16}}
\put(382,40){\vector(1,0){16}}

\put(382,80){\vector(1,0){16}}

\put(398,158){\vector(-1,-1){16}}
\put(398,138){\vector(-1,-1){16}}
\put(398,118){\vector(-1,-1){16}}
\put(398,58){\vector(-1,-1){16}}
\put(398,38){\vector(-1,-1){16}}

{\thicklines\color{blue}
\put(382,62){\vector(1,1){16}}
\put(398,82){\vector(-1,1){16}}
}

\put(388,160){$\vdots$}
\put(388,10){$\vdots$}

{\thicklines\color{red}
\qbezier(402,62)(420,80)(402,98)
\put(404,96){\vector(-1,1){2}}
}

\end{picture}

\begin{picture}(430,180)

\put(10,90){$\stackrel{\mu_2}{\longrightarrow} $}

\put(80,40){\circle{4}}
\put(80,60){\circle{4}}
\put(80,80){\circle{4}}
\put(80,100){\circle{4}}
\put(80,120){\circle{4}}
\put(80,140){\circle{4}} 

\put(100,40){\circle{4}}
\put(100,60){\circle{4}}
\put(100,80){\circle{4}}
\put(100,100){\circle{4}}
\put(100,120){\circle{4}}
\put(100,140){\circle{4}} 

\put(63,135){$\underline{4}$}
\put(63,115){$\underline{3}$}
\put(63,95){$\underline{2}$}
\put(63,75){$\underline{1}$}
\put(63,55){$\underline{n}$}
\put(43,35){$\underline{n-1}$}

\put(110,135){$4$}
\put(110,115){$3$}
\put(125,95){$2$}
\put(125,75){$1$}
\put(110,55){$n$}
\put(110,35){$n-1$}
{\thicklines\color{blue}
\put(80,142){\vector(0,1){16}}
\put(80,122){\vector(0,1){16}}
\put(80,102){\vector(0,1){16}}
\put(80,42){\vector(0,1){16}}
\put(80,22){\vector(0,1){16}}

\put(100,82){\vector(0,1){16}}
}

\put(80,98){\vector(0,-1){16}}

{\thicklines\color{red}
\put(100,142){\vector(0,1){16}}
\put(100,122){\vector(0,1){16}}
\put(100,42){\vector(0,1){16}}
\put(100,22){\vector(0,1){16}}
}

\put(100,118){\vector(0,-1){16}}

\put(82,140){\vector(1,0){16}}
\put(82,120){\vector(1,0){16}}
\put(82,80){\vector(1,0){16}}
\put(82,40){\vector(1,0){16}}

{\thicklines\color{blue}
\put(98,100){\vector(-1,0){16}}
}

\put(98,158){\vector(-1,-1){16}}
\put(98,138){\vector(-1,-1){16}}
\put(98,58){\vector(-1,-1){16}}
\put(98,38){\vector(-1,-1){16}}

{\thicklines\color{blue}
\put(82,62){\vector(1,1){16}}
}
\put(88,160){$\vdots$}
\put(88,10){$\vdots$}

\qbezier(102,62)(120,80)(102,98)
\put(111,80){\vector(0,-1){2}}

{\thicklines\color{red}
\qbezier(102,62)(140,90)(102,118)
\put(104,116){\vector(-1,1){2}}
}

\put(160,90){$\stackrel{\mu_{\underline 2}}{\longrightarrow} $}

\put(230,140){\circle{4}} 
\put(230,120){\circle{4}}
\put(230,100){\circle{4}}
\put(230,80){\circle{4}}
\put(230,60){\circle{4}}
\put(230,40){\circle{4}}

\put(250,140){\circle{4}} 
\put(250,120){\circle{4}}
\put(250,100){\circle{4}}
\put(250,80){\circle{4}}
\put(250,60){\circle{4}}
\put(250,40){\circle{4}}

\put(213,135){$\underline{4}$}
\put(213,115){$\underline{3}$}
\put(213,95){$\underline{2}$}
\put(213,75){$\underline{1}$}
\put(213,55){$\underline{n}$}
\put(193,35){$\underline{n-1}$}

\put(260,135){$4$}
\put(260,115){$3$}
\put(275,95){$2$}
\put(275,75){$1$}
\put(260,55){$n$}
\put(260,35){$n-1$}

{\thicklines\color{blue}
\put(230,142){\vector(0,1){16}}
\put(230,122){\vector(0,1){16}}
\put(230,42){\vector(0,1){16}}
\put(230,22){\vector(0,1){16}}

\put(250,82){\vector(0,1){16}}
}

\put(230,82){\vector(0,1){16}}

\put(230,118){\vector(0,-1){16}}

{\thicklines\color{red}
\put(250,142){\vector(0,1){16}}
\put(250,122){\vector(0,1){16}}
\put(250,42){\vector(0,1){16}}
\put(250,22){\vector(0,1){16}}
}

\put(250,118){\vector(0,-1){16}}

\put(232,140){\vector(1,0){16}}
\put(232,120){\vector(1,0){16}}
\put(232,100){\vector(1,0){16}}
\put(232,80){\vector(1,0){16}}
\put(232,40){\vector(1,0){16}}

\put(248,158){\vector(-1,-1){16}}
\put(248,138){\vector(-1,-1){16}}
\put(248,98){\vector(-1,-1){16}}
\put(248,58){\vector(-1,-1){16}}
\put(248,38){\vector(-1,-1){16}}

{\thicklines\color{blue}
\put(232,62){\vector(1,1){16}}
\put(248,102){\vector(-1,1){16}}
}

\put(238,160){$\vdots$}
\put(238,10){$\vdots$}

\qbezier(252,62)(270,80)(252,98)
\put(261,80){\vector(0,-1){2}}

{\thicklines\color{red}
\qbezier(252,62)(290,90)(252,118)
\put(254,116){\vector(-1,1){2}}
}

\put(310,90){$\stackrel{  \mu_3 }{\longrightarrow} $}


\put(380,140){\circle{4}} 
\put(380,120){\circle{4}}
\put(380,100){\circle{4}}
\put(380,80){\circle{4}}
\put(380,60){\circle{4}}
\put(380,40){\circle{4}}

\put(400,140){\circle{4}} 
\put(400,120){\circle{4}}
\put(400,100){\circle{4}}
\put(400,80){\circle{4}}
\put(400,60){\circle{4}}
\put(400,40){\circle{4}}

\put(363,135){$\underline{4}$}
\put(363,115){$\underline{3}$}
\put(363,95){$\underline{2}$}
\put(363,75){$\underline{1}$}
\put(363,55){$\underline{n}$}
\put(343,35){$\underline{n-1}$}

\put(410,135){$4$}
\put(425,115){$3$}
\put(425,95){$2$}
\put(425,75){$1$}
\put(410,55){$n$}
\put(410,35){$n-1$}

{\thicklines\color{blue}
\put(380,142){\vector(0,1){16}}
\put(380,122){\vector(0,1){16}}
\put(380,42){\vector(0,1){16}}
\put(380,22){\vector(0,1){16}}

\put(400,102){\vector(0,1){16}}
\put(400,82){\vector(0,1){16}}
}

\put(380,82){\vector(0,1){16}}

\put(380,118){\vector(0,-1){16}}

{\thicklines\color{red}
\put(400,142){\vector(0,1){16}}

\put(400,42){\vector(0,1){16}}
\put(400,22){\vector(0,1){16}}
}
\put(400,138){\vector(0,-1){16}}

\put(382,140){\vector(1,0){16}}
\put(382,100){\vector(1,0){16}}
\put(382,80){\vector(1,0){16}}
\put(382,40){\vector(1,0){16}}

{\thicklines\color{blue}
\put(398,120){\vector(-1,0){16}}
}

\put(398,158){\vector(-1,-1){16}}
\put(398,98){\vector(-1,-1){16}}

\put(398,58){\vector(-1,-1){16}}
\put(398,38){\vector(-1,-1){16}}

{\thicklines\color{blue}
\put(382,62){\vector(1,1){16}}
}

\put(388,160){$\vdots$}
\put(388,10){$\vdots$}

\qbezier(402,62)(420,90)(402,118)
\put(411,90){\vector(0,-1){2}}

{\thicklines\color{red}
\qbezier(402,62)(440,100)(402,138)
\put(404,136){\vector(-1,1){2}}
}
\put(388,160){$\vdots$}
\put(388,10){$\vdots$}

\end{picture}

\begin{picture}(430,180)

\put(10,90){$\stackrel{\mu_{\underline 3}}{\longrightarrow} $}

\put(80,40){\circle{4}}
\put(80,60){\circle{4}}
\put(80,80){\circle{4}}
\put(80,100){\circle{4}}
\put(80,120){\circle{4}}
\put(80,140){\circle{4}} 

\put(100,40){\circle{4}}
\put(100,60){\circle{4}}
\put(100,80){\circle{4}}
\put(100,100){\circle{4}}
\put(100,120){\circle{4}}
\put(100,140){\circle{4}} 

\put(63,135){$\underline{4}$}
\put(63,115){$\underline{3}$}
\put(63,95){$\underline{2}$}
\put(63,75){$\underline{1}$}
\put(63,55){$\underline{n}$}
\put(43,35){$\underline{n-1}$}

\put(110,135){$4$}
\put(125,115){$3$}
\put(125,95){$2$}
\put(125,75){$1$}
\put(110,55){$n$}
\put(110,35){$n-1$}

{\thicklines\color{blue}
\put(80,142){\vector(0,1){16}}
\put(80,42){\vector(0,1){16}}
\put(80,22){\vector(0,1){16}}

\put(100,102){\vector(0,1){16}}
\put(100,82){\vector(0,1){16}}
}

\put(80,102){\vector(0,1){16}}
\put(80,82){\vector(0,1){16}}

\put(80,138){\vector(0,-1){16}}

{\thicklines\color{red}
\put(100,142){\vector(0,1){16}}
\put(100,42){\vector(0,1){16}}
\put(100,22){\vector(0,1){16}}
}
\put(100,138){\vector(0,-1){16}}

\put(82,140){\vector(1,0){16}}
\put(82,120){\vector(1,0){16}}
\put(82,100){\vector(1,0){16}}
\put(82,80){\vector(1,0){16}}
\put(82,40){\vector(1,0){16}}

\put(98,158){\vector(-1,-1){16}}
\put(98,118){\vector(-1,-1){16}}
\put(98,98){\vector(-1,-1){16}}

\put(98,58){\vector(-1,-1){16}}
\put(98,38){\vector(-1,-1){16}}

{\thicklines\color{blue}
\put(82,62){\vector(1,1){16}}
\put(98,122){\vector(-1,1){16}}
}

\qbezier(102,62)(120,90)(102,118)
\put(111,90){\vector(0,-1){2}}

{\thicklines\color{red}
\qbezier(102,62)(140,100)(102,138)
\put(104,136){\vector(-1,1){2}}
}

\put(88,160){$\vdots$}
\put(88,10){$\vdots$}

\put(160,90){$\stackrel{\mu_4}{\longrightarrow}$}

\put(233,90){$\cdots$}

\put(310,90){$\stackrel{  \mu_{\underline{n-3}} }{\longrightarrow} $}

\put(380,40){\circle{4}}
\put(380,60){\circle{4}}
\put(380,80){\circle{4}}
\put(380,100){\circle{4}}
\put(380,120){\circle{4}}
\put(380,140){\circle{4}} 

\put(400,40){\circle{4}}
\put(400,60){\circle{4}}
\put(400,80){\circle{4}}
\put(400,100){\circle{4}}
\put(400,120){\circle{4}}
\put(400,140){\circle{4}} 

\put(363,135){$\underline{2}$}
\put(363,115){$\underline{1}$}
\put(363,95){$\underline{n}$}
\put(343,75){$\underline{n-1}$}
\put(343,55){$\underline{n-2}$}
\put(343,35){$\underline{n-3}$}

\put(410,135){$2$}
\put(410,115){$1$}
\put(425,95){$n$}
\put(425,75){$n-1$}
\put(425,55){$n-2$}
\put(410,35){$n-3$}

\put(380,142){\vector(0,1){16}}
\put(380,122){\vector(0,1){16}}
{\thicklines\color{blue}
\put(380,82){\vector(0,1){16}}
\put(380,62){\vector(0,1){16}}
}
\put(380,22){\vector(0,1){16}}

\put(380,58){\vector(0,-1){16}}

{\thicklines\color{blue}
\put(400,142){\vector(0,1){16}}
\put(400,122){\vector(0,1){16}}
\put(400,22){\vector(0,1){16}}
}
\put(400,58){\vector(0,-1){16}}

{\thicklines\color{red}
\put(400,82){\vector(0,1){16}}
\put(400,62){\vector(0,1){16}}
}

\put(382,140){\vector(1,0){16}}
\put(382,120){\vector(1,0){16}}
\put(382,80){\vector(1,0){16}}
\put(382,60){\vector(1,0){16}}
\put(382,40){\vector(1,0){16}}

\put(398,158){\vector(-1,-1){16}}
\put(398,138){\vector(-1,-1){16}}
\put(398,98){\vector(-1,-1){16}}
\put(398,78){\vector(-1,-1){16}}
\put(398,38){\vector(-1,-1){16}}

{\thicklines\color{blue}
\put(398,42){\vector(-1,1){16}}
\put(382,102){\vector(1,1){16}}
}{\thicklines\color{red}
\qbezier(402,62)(420,80)(402,98)
\put(404,64){\vector(-1,-1){2}}
}

\qbezier(402,42)(440,70)(402,98)
\put(421,70){\vector(0,1){2}}

\put(388,160){$\vdots$}
\put(388,10){$\vdots$}

\end{picture}

\begin{picture}(430,180)

\put(10,90){$ \stackrel{\mu_{n-2}}{\longrightarrow} $}

\put(80,140){\circle{4}} 
\put(80,120){\circle{4}}
\put(80,100){\circle{4}}
\put(80,80){\circle{4}}
\put(80,60){\circle{4}}
\put(80,40){\circle{4}}

\put(100,140){\circle{4}} 
\put(100,120){\circle{4}}

{\thicklines\color{red}
\put(100,100){\circle{4}}
\put(100,80){\circle{4}}
}

\put(100,60){\circle{4}}
\put(100,40){\circle{4}}

\put(63,135){$\underline{2}$}
\put(63,115){$\underline{1}$}
\put(63,95){$\underline{n}$}
\put(43,75){$\underline{n-1}$}
\put(43,55){$\underline{n-2}$}
\put(43,35){$\underline{n-3}$}

\put(110,135){$2$}
\put(110,115){$1$}
\put(110,95){$n$}
\put(115,75){$n-1$}
\put(110,55){$n-2$}
\put(110,35){$n-3$}

\put(80,142){\vector(0,1){16}}
\put(80,122){\vector(0,1){16}}
{\thicklines\color{blue}
\put(80,82){\vector(0,1){16}}
\put(80,62){\vector(0,1){16}}
}
\put(80,22){\vector(0,1){16}}

\put(80,58){\vector(0,-1){16}}

{\thicklines\color{blue}
\put(100,142){\vector(0,1){16}}
\put(100,122){\vector(0,1){16}}
\put(100,42){\vector(0,1){16}}
\put(100,22){\vector(0,1){16}}
}
\put(100,78){\vector(0,-1){16}}

\put(82,140){\vector(1,0){16}}
\put(82,120){\vector(1,0){16}}
\put(82,80){\vector(1,0){16}}
\put(82,40){\vector(1,0){16}}

{\thicklines\color{blue}
\put(98,60){\vector(-1,0){16}}
}

\put(98,158){\vector(-1,-1){16}}
\put(98,138){\vector(-1,-1){16}}
\put(98,98){\vector(-1,-1){16}}
\put(98,38){\vector(-1,-1){16}}

{\thicklines\color{blue}
\put(82,102){\vector(1,1){16}}
}

\qbezier(102,62)(120,80)(102,98)
\put(104,96){\vector(-1,1){2}}

\put(88,160){$\vdots$}
\put(88,10){$\vdots$}

\put(160,90){$\stackrel{\mu_{\underline{n-2}}}{\longrightarrow} $}

\put(230,140){\circle{4}} 
\put(230,120){\circle{4}}
\put(230,100){\circle{4}}
\put(230,80){\circle{4}}
\put(230,60){\circle{4}}
\put(230,40){\circle{4}}

\put(250,140){\circle{4}} 
\put(250,120){\circle{4}}

{\thicklines\color{red}
\put(250,100){\circle{4}}
\put(250,80){\circle{4}}
}

\put(250,60){\circle{4}}
\put(250,40){\circle{4}}

\put(213,135){$\underline{2}$}
\put(213,115){$\underline{1}$}
\put(213,95){$\underline{n}$}
\put(193,75){$\underline{n-1}$}
\put(193,55){$\underline{n-2}$}
\put(193,35){$\underline{n-3}$}

\put(260,135){$2$}
\put(260,115){$1$}
\put(260,95){$n$}
\put(265,75){$n-1$}
\put(260,55){$n-2$}
\put(260,35){$n-3$}

\put(230,142){\vector(0,1){16}}
\put(230,122){\vector(0,1){16}}
\put(230,42){\vector(0,1){16}}
{\thicklines\color{blue}
\put(230,82){\vector(0,1){16}}
}
\put(230,22){\vector(0,1){16}}
\put(230,78){\vector(0,-1){16}}

{\thicklines\color{blue}
\put(250,142){\vector(0,1){16}}
\put(250,122){\vector(0,1){16}}
\put(250,42){\vector(0,1){16}}
\put(250,22){\vector(0,1){16}}
}
\put(250,78){\vector(0,-1){16}}

\put(232,140){\vector(1,0){16}}
\put(232,120){\vector(1,0){16}}
\put(232,80){\vector(1,0){16}}
\put(232,60){\vector(1,0){16}}
\put(232,40){\vector(1,0){16}}

\put(248,158){\vector(-1,-1){16}}
\put(248,138){\vector(-1,-1){16}}
\put(248,98){\vector(-1,-1){16}}
\put(248,58){\vector(-1,-1){16}}
\put(248,38){\vector(-1,-1){16}}

{\thicklines\color{blue}
\put(248,62){\vector(-1,1){16}}
\put(232,102){\vector(1,1){16}}
}

\qbezier(252,62)(270,80)(252,98)
\put(254,96){\vector(-1,1){2}}

\put(238,160){$\vdots$}
\put(238,10){$\vdots$}

\put(310,90){$=  M(Q^{(0)})$}

\end{picture}
\end{center}
An intermediate quiver 
$Q^{(2k-1)}=\mu_{k,\underline{k-1}, k-1 , \ldots,\underline{2},2, \underline{1},1}(Q^{(0)})$ 
for $1 \leq k \leq n-2$
contains two balanced cycles
$C^{(2k-1)}=(k+1 \to k+2 \to \cdots \to n \to k+1)$ 
and
${\underline C}^{(2k-1)}=(1 \to 2 \to \cdots \to k \to \underline{k} \to \underline{k+1} \to \cdots \to \underline{n} \to 1 )$ 
of length $n-k$ and $n+1$, respectively.
Similarly, 
$Q^{(2k)}=\mu_{\underline{k}, k , \ldots,\underline{2},2, \underline{1},1}(Q)$ 
 contains two balanced cycles
$C^{(2k)}=C^{(2k-1)}$ 
and 
${\underline C}^{(2k)}=(1 \to 2 \to \cdots \to k \to  \underline{k+1} \to \underline{k+2} \to \cdots \to \underline{n} \to 1 )$ 
of length $n-k$ and $n$, respectively.
Taking into account the rotational symmetry (\ref{eq:rotation})
at each step of mutations,
we find that
\begin{equation} \label{eq:conj}
R_C= M^{-1} \circ R_{C^{(2n-4)}} \circ M \quad
\text{and}  
\quad 
R_{\underline C}= M^{-1} \circ R_{\underline{C}^{(2n-4)}} \circ M.
\end{equation}
In particular,
the resulting quiver $Q^{(2n-4)}=M(Q^{(0)})$
takes the form of two cycles
$C^{(2n-4)}=(n-1 \to n \to n-1)$ and $\underline{C}^{(2n-4)}=(1 \to 2 \to \cdots \to n-2 \to \underline{n-1} \to \underline{n} \to 1) $
of length $2$ and $n$, respectively,
which are connected with a hinge
$(n \to \ul{n-1} \to n-1 \to n-2 \to n)$.
Accordingly, it follows from Theorem~\ref{thm:hinge} that
$(R_{C^{(2n-4)}} \circ R_{\underline{C}^{(2n-4)}})^3 ={\rm id}$,
which is equivalent to $(R_{C} \circ R_{\underline{C}})^3 ={\rm id}$ via
(\ref{eq:conj}).
\qed

\begin{remark}\rm  
Theorems~\ref{thm:hinge} and \ref{thm:ladder}  above can be unified into a more general setting. 
Let $Q$ be a quiver containing two balanced cycles
$C=(1 \to 2 \to \cdots \to n \to 1)$ and 
$\underline{C}=(\underline 1 \to \underline 2 \to \cdots \to \underline m \to  \underline 1 )$
whose vertex sets are  
$I=\{1,2,\ldots,n\}$
and 
$\underline{I}=\{  \underline{1},  \underline{2}, \ldots,  \underline{m}\}$, respectively.
Take subsets $\{u_i \}_{1 \leq i \leq \ell} \subset I$ and $\{ \underline{v_i} \}_{1 \leq i \leq \ell} \subset \underline I$ 
of vertices of the cycles 
so that
$1 \leq u_1 <u_2 < \cdots <u_\ell \leq n$ and $1 \leq v_1 <v_2 < \cdots <v_\ell \leq m$ hold, 
where $\ell \leq \min \{n, m\}$.
Suppose $C$ and $\underline{C}$ are connected with
$\ell$ consecutive wedge graphs $\underline{v_i} \to u_i \to \underline{v_{i-1}}$ $(i \in {\mathbb Z}/\ell {\mathbb Z})$
while there is no other edge between $I$ and $\underline I$.
Then, by applying the mutations at the vertices $I \setminus \{u_i \}_{1 \leq i \leq \ell}$ and 
$\underline{I} \setminus \{ \underline{v_i} \}_{1 \leq i \leq \ell}$ 
not adjacent to each other's cycle,
the quiver $Q$ can be
converted into 
the case of Theorem~\ref{thm:ladder}.
Therefore, the relation
$(R_{C} \circ R_{\underline{C}})^3 ={\rm id}$
is still valid for $Q$.

Note that
(i) if $n=m=\ell$ then $Q$ reduces to the case of Theorem~\ref{thm:ladder};
(ii) if $\ell=2$ and $u_2-u_1=v_2-v_1=1$ then $Q$ reduces to the case of Theorem~\ref{thm:hinge}.
\end{remark}

\section{Examples of birational representations of Weyl groups}
\label{sect:ex}

Starting from the general framework discussed above, 
we can construct 
 an extensive class of birational representations of Weyl groups
 from cluster algebras.
The reflections associated with balanced cycles,
supplemented by appropriate transpositions of vertices, 
generate a Weyl group.
By assembling cycle graphs suitably, we can build the quiver concerned 
while its correspondence with a Dynkin diagram is clearly visible.
In this section we show some examples of affine type,
thereby
relevant to
the $q$-Painlev\'e equations and their higher-order extensions.

\subsection{The $q$-Painlev\'e equation of type 
$D_5^{(1)}$: $q$-$P_{\rm VI}$}
\label{subsec:qp6}

First we consider 
a cycle graph of length four:
\[Q_0=(V_0,E_0), \quad
V_0=\{1,2,3,4\}, \quad
E_0=\{ i \to i+1 \, | \, i \in {\mathbb Z}/4{\mathbb Z}\}.
\]
Notice that $Q_0$ can be regarded as a quiver such that two balanced cycles 
$C_{13}=(1 \to 3 \to 1)$ and $C_{24}=(2 \to 4 \to 2)$
of length two are connected with a hinge $(1 \to 2 \to 3 \to 4 \to 1)$
(see Section~\ref{subsec:hinge}); 
or equivalently $Q_0$ can be regarded as a quiver such that $C_{13}$ and $C_{24}$ are connected with wedge graphs 
$1 \to 2 \to 3$ and $3 \to 4 \to 1$ like a ladder (see Section~\ref{subsec:ladder}).  
Either way, 
we find that
the reflections $R_{13}$ and $R_{24}$ respectively associated with the cycles 
$C_{13}$ and $C_{24}$ keep  $Q_0$ invariant from 
Theorem~\ref{thm:R-inv}, and that they satisfy $(R_{13} \circ R_{24})^3={\rm id}$ from Theorem~\ref{thm:hinge} or ~\ref{thm:ladder}.
Accordingly, the group $G_{Q_0}$ that keeps $Q_0$ invariant includes 
a group  $\langle R_{13}, R_{24}\rangle$ isomorphic to $W(A_2)$,
i.e. the
Weyl group of type $A_2$:
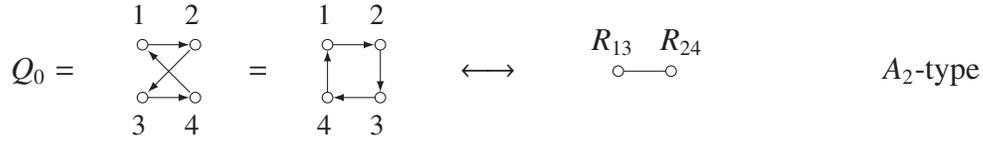
\begin{center}
\begin{picture}(300,50)

\put(-30,17){$Q_0=$}

\put(20,30){\circle{4}}  
\put(16,38){\small$1$} 
\put(22,30){\vector(1,0){16}}
\put(40,30){\circle{4}}  
\put(36,38){\small$2$} 
\put(20,10){\circle{4}}
\put(16,-4){\small$3$} 
\put(22,10){\vector(1,0){16}}
\put(40,10){\circle{4}}  
\put(36,-4){\small$4$} 
\put(38,28){\vector(-1,-1){16}}
\put(38,12){\vector(-1,1){16}}

\put(60,17){$=$}

\put(90,30){\circle{4}}  
\put(86,38){\small$1$} 
\put(92,30){\vector(1,0){16}}
\put(110,30){\circle{4}}  
\put(106,38){\small$2$} 
\put(90,10){\circle{4}}
\put(86,-4){\small$4$} 
\put(108,10){\vector(-1,0){16}}
\put(110,10){\circle{4}}  
\put(106,-4){\small$3$} 
\put(90,12){\vector(0,1){16}}
\put(110,28){\vector(0,-1){16}}

\put(140,17){$\longleftrightarrow$}

\put(200,20){\circle{4}}  
\put(190,28){$R_{13}$} 
\put(202,20){\line(1,0){16}}
\put(220,20){\circle{4}}  
\put(216,28){$R_{24}$} 

\put(300,17){$A_2$-type}
\end{picture}
\end{center}
In the above figure,
we write the quiver on the left and 
the Dynkin diagram corresponding to the Weyl group on the right. 

Next we consider a quiver $Q_1$ obtained from $Q_0$ by adding a copy 
$1'$ of the vertex $1$. 
Obviously, $Q_1$ is invariant under a transposition $(1,1')$ of vertices. Proposition~\ref{prop:copy}
tells us that
$(R_{13} \circ (1,1'))^3={\rm id}$. 
Besides, $(1,1')$ and $R_{24}$ mutually commute.
It thus holds that $G_{Q_1} \supset \langle R_{13}, R_{24}, (1,1') \rangle \simeq W(A_3)$:
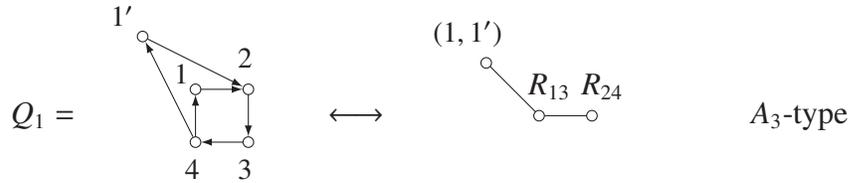
\begin{center}
\begin{picture}(300,60)

\put(20,17){$Q_1=$}
\put(90,30){\circle{4}}  
\put(82,32){\small$1$} 
\put(92,30){\vector(1,0){16}}
\put(110,30){\circle{4}}  
\put(106,38){\small$2$} 
\put(90,10){\circle{4}}
\put(86,-4){\small$4$} 
\put(108,10){\vector(-1,0){16}}
\put(110,10){\circle{4}}  
\put(106,-4){\small$3$} 
\put(90,12){\vector(0,1){16}}
\put(110,28){\vector(0,-1){16}}
\put(58,54){\small$1'$} 
\put(70,50){\circle{4}}  
\put(89,12){\vector(-1,2){18}}
\put(72,49){\vector(2,-1){36}}

\put(140,17){$\longleftrightarrow$}

\put(220,20){\circle{4}}  
\put(216,28){$R_{13}$} 
\put(222,20){\line(1,0){16}}
\put(240,20){\circle{4}}  
\put(236,28){$R_{24}$} 

\put(180,48){\small$(1,1')$} 
\put(200,40){\circle{4}} 
\put(201.5,38.5){\line(1,-1){17}}

\put(300,17){$A_3$-type}
\end{picture}
\end{center}

In the same manner, if we consider a quiver  
\begin{align*}
Q&=(V,E), \\
V&=\{1,2,3,4,1',2',3',4'\}, \quad
E=\{ i \to i+1, \,  i' \to i+1, \, i \to (i+1)', \, i' \to (i+1)' \, | \, i \in {\mathbb Z}/4{\mathbb Z}\}
\end{align*}
obtained from $Q_0$ by adding a copy $i'$ for each vertex $i$
($i \in {\mathbb Z}/4{\mathbb Z}$)
then we observe that
\[
G_Q \supset W= \langle R_{13}, R_{24}, (1,1'),(2,2'), (3,3'),(4,4')\rangle \simeq W(D_5^{(1)});
\]
i.e. the affine Weyl group of type $D^{(1)}_{5}$ naturally emerges:
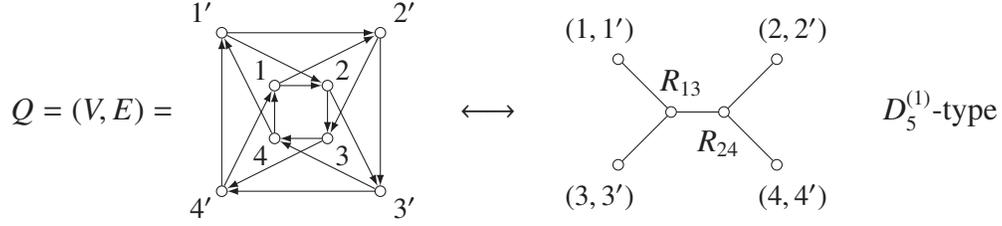
\begin{center}
\begin{picture}(300,80)
\put(-30,37){$Q=(V,E)=$}
\put(70,50){\circle{4}}  
\put(62,52){\small$1$} 
\put(72,50){\vector(1,0){16}}
\put(90,50){\circle{4}}  
\put(93,52){\small$2$} 
\put(70,30){\circle{4}}
\put(62,20){\small$4$} 
\put(88,30){\vector(-1,0){16}}
\put(90,30){\circle{4}}  
\put(93,20){\small$3$} 
\put(70,32){\vector(0,1){16}}
\put(90,48){\vector(0,-1){16}}
\put(38,74){\small$1'$} 
\put(50,70){\circle{4}}  
\put(69,32){\vector(-1,2){18}}
\put(52,69){\vector(2,-1){36}}

\put(115,74){\small$2'$} 
\put(110,70){\circle{4}}  
\put(109,68){\vector(-1,-2){18}}
\put(72,51){\vector(2,1){36}}

\put(52,70){\vector(1,0){56}}  

\put(38,0){\small$4'$} 
\put(50,10){\circle{4}}  
\put(51,12){\vector(1,2){18}}
\put(88,29){\vector(-2,-1){36}}

\put(115,0){\small$3'$} 
\put(110,10){\circle{4}}  
\put(91,48){\vector(1,-2){18}}
\put(108,11){\vector(-2,1){36}}

\put(108,10){\vector(-1,0){56}}  
\put(50,12){\vector(0,1){56}}
\put(110,68){\vector(0,-1){56}}

\put(140,37){$\longleftrightarrow$}

\put(220,40){\circle{4}}  
\put(216,48){$R_{13}$} 
\put(222,40){\line(1,0){16}}
\put(240,40){\circle{4}}  
\put(230,25){$R_{24}$} 

\put(180,68){\small$(1,1')$} 
\put(200,60){\circle{4}} 
\put(201.5,58.5){\line(1,-1){17}}

\put(180,5){\small$(3,3')$} 
\put(200,20){\circle{4}} 
\put(201.5,21.5){\line(1,1){17}}

\put(253,68){\small$(2,2')$} 
\put(260,60){\circle{4}} 
\put(241.5,41.5){\line(1,1){17}}

\put(253,5){\small$(4,4')$} 
\put(260,20){\circle{4}} 
\put(241.5,38.5){\line(1,-1){17}}

\put(300,37){$D_5^{(1)}$-type}
\end{picture}
\end{center}

By means of  Proposition~\ref{prop:birat},
the birational transformations of the generators
\[
s_0=(1,1'), \quad s_1= (3,3'), \quad s_2=R_{13}, \quad 
s_3=R_{24}, \quad s_4=(4,4'), \quad s_5=(2,2')
\]
of $W(D_5^{(1)})$
on the variables 
$y_i$ $(i=1,2,3,4,1',2',3',4')$
attached to the vertices of $Q$
are described as follows:
\begin{equation}   \label{eq:D5}
\begin{aligned}
&s_0:y_1 \leftrightarrow y_{1'}, \quad s_1:y_3 \leftrightarrow y_{3'}, \quad s_4:y_4 \leftrightarrow y_{4'}, \quad s_5:y_2 \leftrightarrow y_{2'}, 
\\ 
&s_2(y_{\{1,3\}})= \frac{1}{y_{\{3,1\}}}, \quad
s_2(y_{\{2,2'\}})= y_{\{2,2'\}} \frac{y_1(1+y_3)}{1+y_1},\quad
s_2(y_{\{4,4'\}})= y_{\{4,4'\}}  \frac{y_3(1+y_1)}{1+y_3},
\\
&s_3(y_{\{2,4\}})= \frac{1}{y_{\{4,2\}}}, \quad
s_3(y_{\{1,1'\}})= y_{\{1,1'\}} \frac{y_4(1+y_2)}{1+y_4},\quad
s_3(y_{\{3,3'\}})= y_{\{3,3'\}} \frac{y_2(1+y_4)}{1+y_2}.
\end{aligned}
\end{equation}
Here we have omitted to write the action on the variables if it is trivial.
The compositions of permutations and the inversion $\iota$ defined by
\begin{equation}   \label{eq:D5_auto}
\begin{aligned}
&\sigma_1=(1,2)\circ(1',2')\circ(3,4)\circ(3',4')\circ \iota :
\quad
y_{\{1,1',3,3'\}} \leftrightarrow \frac{1}{y_{\{2, 2', 4, 4' \}}}, 
\\
&\sigma_2=(1,3) \circ(1',3')\circ \iota : 
\quad
y_{\{1,1'\}} \leftrightarrow \frac{1}{y_{\{3, 3'\}}}, 
\quad y_{\{2,2',4,4'\}} \mapsto \frac{1}{ y_{\{2,2',4,4'\}}  }
\end{aligned}
\end{equation}
also keep $Q$ invariant, i.e. $\langle \sigma_1,\sigma_2\rangle \subset G_Q$,
and represent the Dynkin diagram automorphisms. 
We have the relations 
\[
{s_i}^2={\rm id}, \quad s_i s_j = s_js_i \quad (\text{if $c_{ij}=0$}),
\quad s_i s_j s_i=s_j s_i s_j \quad  (\text{if $c_{ij}=-1$})
\]
and 
\[
{\sigma_1}^2={\sigma_2}^2={\rm id}, \quad
\sigma_1 \circ  s_{\{0,1,2,3,4,5\}}= s_{\{5,4,3,2,1,0\}}\circ \sigma_1,
\quad
\sigma_2 \circ s_{\{0,1\}}=s_{\{1,0\}} \circ \sigma_2,
\]
where $(c_{ij})_{0 \leq i,j \leq 5}$ denotes the 
Cartan matrix of type $D_5^{(1)}$:
\[(c_{ij})_{0 \leq i,j \leq 5} =
\begin{pmatrix}
2&&-1&&&\\
&2&-1&&&\\
-1&-1&2&-1&&\\
&&-1&2&-1&-1\\
&&&-1&2&\\
&&&-1&&2
\end{pmatrix}
\qquad \qquad \qquad \qquad
\begin{picture}(100,40)
\put(20,0){\circle{4}}  
\put(16,8){$2$} 
\put(22,0){\line(1,0){16}}
\put(40,0){\circle{4}}  
\put(35,-15){$3$} 

\put(-5,28){\small$0$} 
\put(0,20){\circle{4}} 
\put(1.5,18.5){\line(1,-1){17}}

\put(-5,-35){\small$1$} 
\put(0,-20){\circle{4}} 
\put(1.5,-18.5){\line(1,1){17}}

\put(58,28){\small$5$} 
\put(60,20){\circle{4}} 
\put(41.5,1.5){\line(1,1){17}}

\put(58,-35){\small$4$} 
\put(60,-20){\circle{4}} 
\put(41.5,-1.5){\line(1,-1){17}}
\end{picture} 
\]

This birational realization (\ref{eq:D5}) and (\ref{eq:D5_auto}) of the {\it extended} affine Weyl group 
$\widetilde{W}(D_5^{(1)})=\langle s_i \ (0 \leq i \leq 5) \rangle \rtimes \langle
\sigma_1, \sigma_2 \rangle$
is equivalent to that arising from 
Cremona isometries of a certain rational surface; cf. \cite{Looijenga, Sakai1}. 
The birational action of a translation $T=(\sigma_1 \sigma_2 s_2 s_0s_1s_2)^2 \in \widetilde{W}(D_5^{(1)})$
reduces to 
a non-autonomous system of $q$-difference equations,
called the {\it sixth $q$-Painlev\'e equation} ($q$-$P_{\rm VI}$).
We shall later 
address the problem of how to find a good coordinate system; see (\ref{eq:qp6}) in Section~\ref{sect:symp}.

\begin{remark}\rm
\label{remark:D5}
In a similar manner, we can construct birational representations of Weyl groups corresponding to various types of Dynkin diagrams.
In this subsection we have started with a quiver $Q_0=(1 \to 2 \to 3 \to 4 \to 1)$, 
which is viewed as two cycles of length two connected with a hinge, 
and then obtained $Q$ by 
adding 
one copy for each vertex of $Q_0$; the resulting Weyl group is of type $D_5^{(1)}$.
Alternatively, 
if we add any number of copy vertices to the same quiver $Q_0$,
then we can reproduce the birational representations of Weyl groups 
acting on certain rational surfaces found in \cite{Tsuda1}. 
That is,
by adding $k_i$ copies for each vertex $i$ ($i=1,2,3,4$) to $Q_0$,  
we are led to
the Weyl group 
corresponding to
the H-shaped Dynkin diagram:
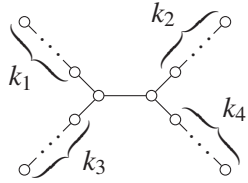
\begin{center}
\begin{picture}(60,70)
\put(-8,58){\circle{4}}   
\put(-6.5,56.5){\line(1,-1){3}}
\put(-4,50.5){\rotatebox{-45}{\footnotesize$\cdots$}}
\put(9.5,40.5){\line(-1,1){3}}
\put(11,39){\circle{4}} 
\put(12.5,37.5){\line(1,-1){6}}
\put(-17,58){\rotatebox{-45}{\footnotesize$\underbrace{\quad \qquad}$}}
\put(-14,34){\small$k_1$}
\put(40,61){\rotatebox{-135}{\footnotesize$\underbrace{\quad \qquad}$}}
\put(40,58){\small$k_2$}
\put(51,3){\rotatebox{135}{\footnotesize$\underbrace{\quad \qquad}$}}
\put(67,22){\small$k_4$}
\put(-6,0){\rotatebox{45}{\footnotesize$\underbrace{\quad \qquad}$}}
\put(14,2){\small$k_3$}
\put(20,30){\circle{4}}   
\put(22,30){\line(1,0){16}}
\put(40,30){\circle{4}} 
\put(12.5,22.5){\line(1,1){6}}
\put(11,21){\circle{4}}
\put(9.5,19.5){\line(-1,-1){3}}
\put(-2.5,6){\rotatebox{45}{\footnotesize$\cdots$}}
\put(-6.5,3.5){\line(1,1){3}}
\put(-8,2){\circle{4}}
\put(41.5,31.5){\line(1,1){6}}
\put(49,39){\circle{4}} 
\put(50.5,40.5){\line(1,1){3}}
\put(54.5,42.9){\rotatebox{45}{\footnotesize$\cdots$}}
\put(66.5,56.5){\line(-1,-1){3}}
\put(68,58){\circle{4}} 
\put(41.5,28.5){\line(1,-1){6}}
\put(49,21){\circle{4}}
\put(50.5,19.5){\line(1,-1){3}}
\put(52.5,13.7){\rotatebox{-45}{\footnotesize$\cdots$}}
\put(66.5,3.5){\line(-1,1){3}}
\put(68,2){\circle{4}}
\end{picture} 
\end{center}
specified by a quartet $(k_1,k_2,k_3,k_4) \in ({\mathbb Z}_{\geq 0})^4$
of nonnegative integers.
For instance,
if we choose $(k_1,k_2,k_3,k_4)=(5,0,2,0)$ then the resulting Weyl group is of type $E_8^{(1)}$,
from which we can literally derive the $q$-Painlev\'e equation of type $E_8^{(1)}$.

Furthermore, if we start with a quiver such that any number of $2$-cycles
are connected in a row with hinges and add any number of copies for each vertex, then we can reproduce
the birational representations of Weyl groups 
acting on certain rational varieties found in \cite{TT};
the corresponding Dynkin diagram is the comb-shaped one $T^{\boldsymbol k}_{\boldsymbol \ell}$ (see Section~\ref{sect:intro}). 
\end{remark}

\subsection{The $q$-Painlev\'e equation of type 
$A_4^{(1)}$}
\label{subsec:qp5}

First we consider a quiver
\begin{align*}
Q_0&=(V_0,E_0), \\
V_0&=\{1,2,3,4,5\}, \quad
E_0=\{ i \to i+1 \, (i=1,2,3), \,  4 \to 1, \, 5 \to j \, (j=1,4), \,  k \to 5 \, (k=2,3) \}
\end{align*}
in which two balanced cycles $C_{13}=(1 \to 3 \to 1)$ and $C_{24}=(2 \to 4 \to 2)$
of length two are adjacent in a ladder shape and 
both of them intersect with 
another balanced cycle 
$C_{125}=(1 \to 2 \to 5 \to 1)$
of length three.
By virtue of Theorem~\ref{thm:R-inv},
the reflections $R_{13}$, $R_{24}$ and $R_{125}$ respectively associated with the cycles 
$C_{13}$,  $C_{24}$ and $C_{125}$ keep $Q_0$ invariant.
The relation $(R_{13} \circ R_{24})^3={\rm id}$ holds by Theorem~\ref{thm:hinge} or \ref{thm:ladder}
and both $R_{13}$ and $R_{24}$ commute with $R_{125}$
 by Theorem~\ref{thm:commute}. 
Hence we observe that
\[
G_{Q_0} \supset W_0= \langle R_{13},R_{24}, R_{125}\rangle
\simeq W(A_2) \times W(A_1)
\]
and the correspondence between the quiver and the Dynkin diagram reads as follows:
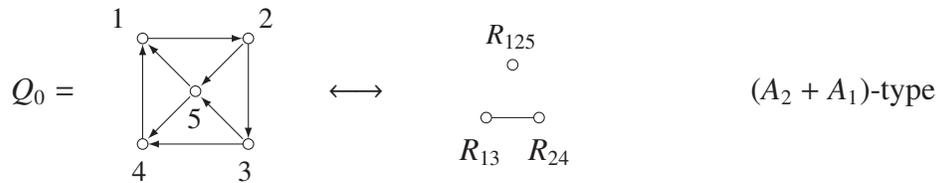
\begin{center}
\begin{picture}(300,60)

\put(20,27){$Q_0=$}
\put(90,30){\circle{4}}  
\put(87,16){\small$5$} 

\put(70,12){\vector(0,1){36}}
\put(110,10){\circle{4}}
\put(66,-4){\small$4$} 

\put(108,10){\vector(-1,0){36}}
\put(70,10){\circle{4}}  
\put(106,-4){\small$3$} 
\put(110,48){\vector(0,-1){36}}
\put(58,54){\small$1$} 
\put(70,50){\circle{4}}  
\put(72,50){\vector(1,0){36}}
\put(114,54){\small$2$} 
\put(110,50){\circle{4}}  

\put(108,48){\vector(-1,-1){16}}
\put(88,28){\vector(-1,-1){16}}
\put(88,32){\vector(-1,1){16}}
\put(108,12){\vector(-1,1){16}}

\put(140,27){$\longleftrightarrow$}

\put(200,48){\small$R_{125}$} 
\put(210,40){\circle{4}} 

\put(200,20){\circle{4}}  
\put(190,4){$R_{13}$} 
\put(202,20){\line(1,0){16}}
\put(220,20){\circle{4}}  
\put(216,4){$R_{24}$} 

\put(300,27){$(A_2+A_1)$-type}
\end{picture}
\end{center}

Next we consider a quiver $Q$ obtained from $Q_0$ by adding copies $1'$ and $2'$ of the vertices  $1$ and $2$, respectively.
Obviously, $Q$ is invariant under transpositions $(1,1')$ and $(2,2')$ of vertices.
Proposition~\ref{prop:copy} shows that  $R_{13} \circ (1,1')$, $R_{24} \circ (2,2')$, $R_{125} \circ (1,1')$ and $R_{125} \circ (2,2')$ are of order three.
Consequently, we observe that
\[
G_{Q} \supset W= \langle R_{13},R_{24}, R_{125},  (1,1'), (2,2')\rangle
\simeq W(A_4^{(1)}).
\]
\begin{center}
\begin{picture}(300,80)

\put(-20,37){$Q=$}
\put(70,30){\circle{4}}  
\put(67,16){\small$5$} 

\put(50,12){\vector(0,1){56}}
\put(90,10){\circle{4}}
\put(46,-4){\small$4$} 

\put(49,12){\vector(-1,2){18}}
\put(109,48){\vector(-1,-2){18}}

\put(69,32){\vector(-1,2){18}}
\put(89,68){\vector(-1,-2){18}}

\put(108,49){\vector(-2,-1){36}}
\put(68,31){\vector(-2,1){36}}

\put(88,10){\vector(-1,0){36}}
\put(50,10){\circle{4}}  
\put(86,-4){\small$3$} 
\put(90,68){\vector(0,-1){56}}
\put(38,74){\small$1'$} 
\put(50,70){\circle{4}}  
\put(52,70){\vector(1,0){36}}
\put(94,74){\small$2'$} 
\put(90,70){\circle{4}} 

\put(18,54){\small$1$} 
\put(30,50){\circle{4}}  
\put(32,50){\vector(1,0){76}}
\put(114,54){\small$2$} 
\put(110,50){\circle{4}}  

\put(68,28){\vector(-1,-1){16}}
\put(88,12){\vector(-1,1){16}}
\put(33,51){\vector(3,1){54}}
\put(53,69){\vector(3,-1){54}}

\put(140,37){$\longleftrightarrow$}

\put(220,64){\small$R_{125}$} 
\put(230,56){\circle{4}} 

\put(210,40){\circle{4}}
\put(178,40){\small$(1,1')$} 
\put(250,40){\circle{4}} 
\put(256,40){\small$(2,2')$} 

\put(220,20){\circle{4}}  
\put(210,4){$R_{13}$} 
\put(222,20){\line(1,0){16}}
\put(240,20){\circle{4}}  
\put(236,4){$R_{24}$} 

\put(219,22){\line(-1,2){8}}
\put(241,22){\line(1,2){8}}

\put(228,55){\line(-5,-4){16.5}}
\put(232,55){\line(5,-4){16.5}}

\put(310,37){$A_4^{(1)}$-type}
\end{picture}
\end{center}

By means of  Proposition~\ref{prop:birat},
the birational transformations of the generators
\[s_0=R_{125}, \quad s_1= (1,1'), \quad s_2=R_{13}, \quad 
s_3=R_{24}, \quad s_4=(2,2')
\]
of $W(A_4^{(1)})$
on the variables 
$y_i$ $(i=1,2,3,4,5,1',2')$
attached to the vertices of $Q$
are described as follows:
\begin{equation}   \label{eq:A4}
\begin{aligned}
&s_1:y_1 \leftrightarrow y_{1'}, \quad  s_4:y_2 \leftrightarrow y_{2'}, 
\\ 
&s_0(y_1)=\frac{1+y_1 +y_1 y_2}{y_2(1+y_5+y_5y_1)},
\quad 
s_0(y_2)=\frac{1+y_2 +y_2 y_5}{y_5(1+y_1+y_1y_2)},
\quad 
s_0(y_5)=\frac{1+y_5 +y_5 y_1}{y_1(1+y_2+y_2y_5)},
\\
&s_0(y_3)=y_3\frac{y_1y_2(1+y_5 +y_5 y_1)}{1+y_1+y_1y_2},
\quad
s_0(y_4)=y_4\frac{y_2y_5(1+y_1 +y_1 y_2)}{1+y_2+y_2y_5},
\\
&s_0(y_{1'})=y_{1'}\frac{y_5(1+y_1 +y_1 y_2)}{1+y_5+y_5y_1},
\quad
s_0(y_{2'})=y_{2'}\frac{y_1(1+y_2 +y_2 y_5)}{1+y_1+y_1y_2},
\\
&s_2(y_{\{1,3\}})= \frac{1}{y_{\{3,1\}}}, \quad
s_2(y_{\{2,2'\}})= y_{\{2,2'\}} \frac{y_1(1+y_3)}{1+y_1},\quad
s_2(y_{\{4,5\}})= y_{\{4,5\}}  \frac{y_3(1+y_1)}{1+y_3},
\\
&s_3(y_{\{2,4\}})= \frac{1}{y_{\{4,2\}}}, \quad
s_3(y_{\{1,1'\}})= y_{\{1,1'\}} \frac{y_4(1+y_2)}{1+y_4},\quad
s_3(y_{\{3,5\}})= y_{\{3,5\}} \frac{y_2(1+y_4)}{1+y_2}.
\end{aligned}
\end{equation}
Here we have omitted to write the action on the variables if it is trivial.
The compositions of permutations and the inversion $\iota$ defined by
\[\sigma_1=(1,2)\circ(1',2')\circ(3,4)\circ \iota 
\quad \text{and}  \quad
\sigma_2=(1,2)\circ(1',2',3,5,4) \circ \mu_2
\]
also keep $Q$ invariant, i.e. $\langle \sigma_1,\sigma_2\rangle \subset G_Q$,  and represent the Dynkin diagram automorphisms.
Their birational actions are given as
\begin{equation}   \label{eq:A4_auto}
\begin{aligned}
&\sigma_1
(y_{\{1,2,3,4,5,1',2'\}}) = \frac{1}{y_{\{2, 1, 4,3,5,2',1' \}}}, 
\\
&
\sigma_2(y_1)=\frac{1}{y_2},
\quad 
\sigma_2(y_{\{2,2'\}})=y_{\{1,1'\}}(1+y_2),
\quad
\sigma_2(y_{\{3,1'\}})= y_{\{2',4\}} ,
\quad
\sigma_2(y_{\{4,5\}}) =\frac{ y_{\{5,3\} }}{1+{y_2}^{-1}}.
\end{aligned}
\end{equation}
We have the relations 
\[
{s_i}^2={\rm id}, \quad s_i s_j = s_js_i \quad (\text{if $c_{ij}=0$}),
\quad s_i s_j s_i=s_j s_i s_j \quad  (\text{if $c_{ij}=-1$})
\]
and 
\[
{\sigma_1}^2={\sigma_2}^5={\rm id}, \quad
\sigma_1 \circ  s_{i}= s_{5-i}\circ \sigma_1, \quad
\sigma_2 \circ s_{i}=s_{i+2} \circ \sigma_2  \quad(i \in {\mathbb Z}/5{\mathbb Z})
\]
where $(c_{ij})_{0 \leq i,j \leq 4}$ denotes the Cartan matrix 
of type $A_4^{(1)}$:
\[(c_{ij})_{0 \leq i,j \leq 4} =
\begin{pmatrix}
2&-1&&&-1\\
-1&2&-1&&\\
&-1&2&-1&\\
&&-1&2&-1\\
-1&&&-1&2
\end{pmatrix}
\qquad \qquad 
\begin{picture}(100,30)
\put(47,28){\small$0$} 
\put(50,21){\circle{4}} 

\put(30,5){\circle{4}}
\put(18,5){\small$1$} 
\put(70,5){\circle{4}} 
\put(76,5){\small$4$} 

\put(40,-15){\circle{4}}  
\put(34,-30){$2$} 
\put(42,-15){\line(1,0){16}}
\put(60,-15){\circle{4}}  
\put(60,-30){$3$} 

\put(39,-13){\line(-1,2){8}}
\put(61,-13){\line(1,2){8}}


\put(48,20){\line(-5,-4){16.5}}
\put(52,20){\line(5,-4){16.5}}
\end{picture}
\]

This birational realization  (\ref{eq:A4}) and (\ref{eq:A4_auto}) of the {\it extended} affine Weyl group 
$\widetilde{W}(A_4^{(1)})=\langle s_i \ (0 \leq i \leq 4) \rangle \rtimes \langle
\sigma_1, \sigma_2 \rangle$
is equivalent to that given in \cite{Sakai1} and its lattice part yields a 
$q$-analogue of the fifth Painlev\'e equation.

\begin{remark}\rm
We may also consider a quiver 
obtained from $Q_0$ by adding $k_i$ copies for each vertex $i$ $(i=1,2, \ldots, 5)$
where $k_i $ is an arbitrary nonnegative integer.
The corresponding 
Dynkin diagram becomes
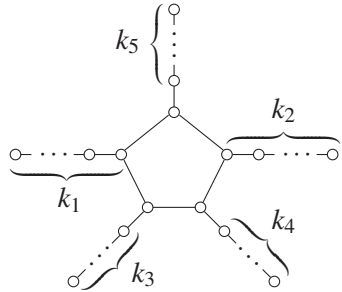
\begin{center}
\begin{picture}(60,110)

\put(30,66){\circle{4}} 
\put(30,68){\line(0,1){8}}
\put(30,78){\circle{4}}
\put(30,80){\line(0,1){4}}
\put(29.5,86){\rotatebox{90}{\footnotesize$\cdots$}}
\put(30,99){\line(0,1){4}}
\put(30,105){\circle{4}} 

\put(20,107){\rotatebox{-90}{\footnotesize$\underbrace{\quad \qquad}$}}
\put(8,90){\small$k_5$}
\put(19,32){\line(-1,2){8}}
\put(41,32){\line(1,2){8}}

\put(28,65){\line(-5,-4){16.5}}
\put(32,65){\line(5,-4){16.5}}
\put(10,50){\circle{4}}
\put(8,50){\line(-1,0){8}}
\put(-2,50){\circle{4}}
\put(-4,50){\line(-1,0){4}}
\put(-21,47.5){\footnotesize$\cdots$}
\put(-24,50){\line(-1,0){4}}
\put(-30,50){\circle{4}}

\put(-32,46){\footnotesize$\underbrace{ \ \qquad \qquad}$}
\put(-14,30){\small$k_1$}
\put(50,50){\circle{4}} 
\put(52,50){\line(1,0){8}}
\put(62,50){\circle{4}}
\put(64,50){\line(1,0){4}}
\put(71,47.5){\footnotesize$\cdots$}
\put(88,50){\line(-1,0){4}}
\put(90,50){\circle{4}}

\put(50,53){\footnotesize$\overbrace{ \ \qquad \qquad}$}
\put(67,64){\small$k_2$}
\put(12.5,22.5){\line(1,1){6}}
\put(11,21){\circle{4}}
\put(9.5,19.5){\line(-1,-1){3}}
\put(-2.5,6){\rotatebox{45}{\footnotesize$\cdots$}}
\put(-6.5,3.5){\line(1,1){3}}
\put(-8,2){\circle{4}}

\put(51,3){\rotatebox{135}{\footnotesize$\underbrace{\quad \qquad}$}}
\put(67,22){\small$k_4$}
\put(-6,0){\rotatebox{45}{\footnotesize$\underbrace{\quad \qquad}$}}
\put(14,2){\small$k_3$}
\put(20,30){\circle{4}}   
\put(22,30){\line(1,0){16}}
\put(40,30){\circle{4}} 

\put(41.5,28.5){\line(1,-1){6}}
\put(49,21){\circle{4}}
\put(50.5,19.5){\line(1,-1){3}}
\put(52.5,13.7){\rotatebox{-45}{\footnotesize$\cdots$}}
\put(66.5,3.5){\line(-1,1){3}}
\put(68,2){\circle{4}}
\end{picture} 
\end{center}
which occurs in Looijenga's work \cite{Looijenga}
related to certain rational surfaces.
\end{remark}

\subsection{An example of higher-dimensional cases}  
 \label{subsec:ex2}

Let $m$ and $n$ be integers greater than one
with at least one of them greater than two.
Consider a toroidal quiver $Q$ consisting of a set of vertices 
$V=\{v_{i,j} \, | \, i \in {\mathbb Z}/m{\mathbb Z}, j \in  {\mathbb Z}/n{\mathbb Z} \}$ and a set of edges $E=\{ v_{i,j} \to v_{i+1,j} , \, v_{i,j} \to v_{i,j+1} , \, v_{i+1,j+1} \to v_{i,j}    \}$:
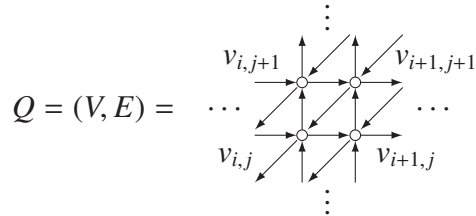
\begin{center}
\begin{picture}(150,100)

\put(-30,47){$Q=(V,E)=$}

\put(80,40){\circle{4}}
\put(80,60){\circle{4}}

\put(100,40){\circle{4}}
\put(100,60){\circle{4}}

\put(50,67){$v_{i,j+1}$}
\put(48,30){$v_{i,j}$}

\put(115,67){$v_{i+1,j+1}$}
\put(108,30){$v_{i+1,j}$}

\put(80,22){\vector(0,1){16}}
\put(80,42){\vector(0,1){16}}
\put(80,62){\vector(0,1){16}}

\put(100,22){\vector(0,1){16}}
\put(100,42){\vector(0,1){16}}
\put(100,62){\vector(0,1){16}}

\put(62,60){\vector(1,0){16}}
\put(62,40){\vector(1,0){16}}

\put(82,60){\vector(1,0){16}}
\put(82,40){\vector(1,0){16}}

\put(102,60){\vector(1,0){16}}
\put(102,40){\vector(1,0){16}}
\put(78,58){\vector(-1,-1){16}}
\put(78,38){\vector(-1,-1){16}}

\put(98,78){\vector(-1,-1){16}}
\put(98,58){\vector(-1,-1){16}}
\put(98,38){\vector(-1,-1){16}}

\put(118,78){\vector(-1,-1){16}}
\put(118,58){\vector(-1,-1){16}}

\put(88,80){$\vdots$}
\put(88,10){$\vdots$}

\put(123,47){$\cdots$}
\put(44,47){$\cdots$}

\end{picture}
\end{center}
Note that omitting the horizontal periodicity reduces $Q$ to the cylindrical quiver discussed in \cite{ILP}.
The quiver 
$Q$ contains {\it vertical} cycles
$C^v_i =(v_{i,1} \to v_{i,2} \to \cdots \to v_{i,n} \to v_{i,1})$ $(i \in  {\mathbb Z}/m{\mathbb Z})$ of length $n$ and {\it horizontal} cycles 
$C^h_j=(v_{1,j} \to v_{2,j} \to \cdots \to v_{m,j} \to v_{1,j})$ $(j \in  {\mathbb Z}/n{\mathbb Z})$ of length $m$
as subgraphs.
In addition, if $m$ and $ n$ are not relatively prime then
$Q$ contains 
{\it diagonal} cycles 
$C^d_k=(v_{k,0} \to v_{k-1,-1} \to v_{k-2,-2} \to \cdots \to v_{k-\ell+1,-\ell+1} \to v_{k,0})$
$(k \in  {\mathbb Z}/g{\mathbb Z})$
of length $\ell =mn/g$,
where $g$ is the greatest common divisor of $m$ and $n$.
Observe that these three types of cycles are all balanced.
Therefore the reflections  
\[ s_i^{\triangle}=R_{C^\triangle_i }, \quad \triangle = v, h, d \]
associated with the cycles 
$C^\triangle_i $  
keep $Q$ invariant according to Theorem~\ref{thm:R-inv}.

For each $\triangle=v,h,d$,
two cycles  $C^\triangle_i$ and $C^\triangle_j$ are adjacent in a ladder shape 
if $|i-j| \equiv1$ and are not connected by any edge if $|i-j| \nequiv 0,1$.
Two cycles $C^\triangle_i$ and  $C^{\triangle'}_j$ intersect each other for any 
$i$ and $j$ 
if $\triangle \neq \triangle'$. 
Theorems~\ref{thm:commute} and \ref{thm:ladder} imply that
the reflections $s_i^{\triangle}$ for $\triangle=v,h,d$
generate groups isomorphic to the affine Weyl groups $W(A^{(1)}_r)$
with $r=m-1,n-1,g-1$, respectively, 
and any two of the three mutually commute;
i.e. the relations
\begin{equation}
\label{eq:rel_A}
(s_i^{\triangle})^2={\rm id} , \quad 
s_i^{\triangle}s_{i+1}^{\triangle}s_i^{\triangle}=
s_{i+1}^{\triangle}s_i^{\triangle}s_{i+1}^{\triangle},
\quad
s_i^{\triangle}s_j^{\triangle}=s_j^{\triangle}s_i^{\triangle}
\quad (|i-j|\nequiv 0,1)
\end{equation}
hold for each $\triangle=v,h,d$,
and
the commutativity 
$s_i^{\triangle}s_{j}^{\triangle'}=s_{j}^{\triangle'}s_i^{\triangle}$
holds
for any 
$i$ and $j$ if $\triangle \neq \triangle'$.
Note that when $m$, $n$ and $g$ equal two,
the second and third relations in (\ref{eq:rel_A}) are omitted for $\triangle=v, h, d$, respectively.
As a consequence, we find that
\[
G_Q \supset W= \langle s_i^{v},  s_i^{h},  s_i^{d}   \rangle
\simeq 
W(A^{(1)}_{m-1})\times 
W(A^{(1)}_{n-1})\times W(A^{(1)}_{g-1}).
\]

By means of Proposition~\ref{prop:birat},
the birational transformations of the generators $s_i^\triangle$
of $W(A^{(1)}_{m-1})\times 
W(A^{(1)}_{n-1})\times W(A^{(1)}_{g-1})$
on the variables $y_{i,j}$  $(i \in {\mathbb Z}/m{\mathbb Z}, j \in  {\mathbb Z}/n{\mathbb Z})$ 
attached to the vertices $v_{i,j}$ of $Q$ are described as follows:
\begin{align}
\label{eq:sv}
s_{i}^v (y_{i,j})&=\frac{F_{i,j-1}}{y_{i,j+1} F_{i,j+1}}, 
\quad 
\frac{s_{i}^v (y_{i+1,j})}{y_{i+1,j}}
=\left\{
\begin{array}{lll}
\displaystyle
\frac{s_{i}^v (y_{i-1,j-1})}{y_{i-1,j-1}}
=
\frac{y_{i,j} F_{i,j}}{F_{i,j-1}}
&(\text{if $m \geq 3$})
\\ 
\\
\displaystyle
\frac{y_{i,j} y_{i,j+1}F_{i,j+1}}{ F_{i,j-1}}
&(\text{if $m = 2$})
\end{array}
\right.
\\
\label{eq:sh} 
s_{j}^h (y_{i,j})&=\frac{G_{i-1,j}}{y_{i+1,j} G_{i+1,j}},
\quad
\frac{s_{j}^h (y_{i,j+1})}{y_{i,j+1} }
=
\left\{
\begin{array}{lll}
\displaystyle
\frac{s_{j}^h (y_{i-1,j-1})}{y_{i-1,j-1} }=
\frac{y_{i,j} G_{i,j}}{G_{i-1,j}}
&(\text{if $n \geq 3$})
\\ 
\\
\displaystyle
\frac{y_{i,j} y_{i+1,j}G_{i+1,j}}{ G_{i-1,j}}
&(\text{if $n = 2$})
\end{array}
\right.
\\
\label{eq:sd}
s_{k}^d (y_{i+k,i})&=\frac{H_{i+k+1,i+1}}{y_{i+k-1,i-1} H_{i+k-1,i-1}}, 
\quad
\frac{s_{k}^d (y_{i+k,i+1})}{y_{i+k,i+1} }
=
\left\{
\begin{array}{lll}
\displaystyle
\frac{s_{k}^d (y_{i+k+1,i})}{y_{i+k+1,i} }
=\frac{y_{i+k,i}  H_{i+k,i}}{H_{i+k+1,i+1}}
&(\text{if $g \geq 3$})
\\
\\
\displaystyle
\frac{y_{i+k,i} y_{i+k-1,i+1} H_{i+k,i} H_{i+k-1,i+1} }{ H_{i+k+1,i+1} H_{i+k,i+2} }
&(\text{if $g =2$})
\end{array}
\right.
\end{align}
where $F_{i,j}$, $G_{i,j}$ and $H_{i,j}$ are polynomials in $y$-variables defined by 
\[
F_{i,j}=1+\sum_{a=1}^{n-1} \prod_{b=1}^{a} y_{i,j+b},
\quad
G_{i,j}=1+\sum_{a=1}^{m-1} \prod_{b=1}^{a} y_{i+b,j},
\quad
H_{i,j}=1+\sum_{a=1}^{mn/g-1} \prod_{b=1}^{a} y_{i-b,j-b}.
\]
We have omitted to write the action on the variables if it is trivial.

Let us summarize the above as a theorem.

\begin{thm} 
The birational transformations {\rm(\ref{eq:sv})--(\ref{eq:sd})} realize
the affine Weyl group  $W(A^{(1)}_{m-1})\times 
W(A^{(1)}_{n-1})\times W(A^{(1)}_{g-1})$
over the field 
${\mathbb Q}(\{y_{i,j} \})$
of rational functions.
\end{thm}

\begin{remark}\label{remark:AA}
\rm
The first two parts of the above formulae, 
(\ref{eq:sv}) and (\ref{eq:sh}),
are equivalent to 
the representation of $W(A^{(1)}_{m-1})\times 
W(A^{(1)}_{n-1})$ due to Yamada \cite{Yamada} and 
Kajiwara--Noumi--Yamada \cite{KNY}
through
a certain change of variables.
The emergence of the reflections (\ref{eq:sd}) associated 
with {\it diagonal} cycles
is thought of as an advantage of the usage of cluster algebras.
The lattice part of the affine Weyl group yields 
$q$-analogues of
the fourth and fifth Painlev\'e equations and their higher-order extensions as mentioned in \cite{KNY}.
Interestingly enough, it was reported by Okubo--Suzuki \cite{OS} that
various kinds of higher-order extensions of 
$q$-$P_{\rm VI}$
including the $q$-Garnier system 
(cf. \cite{Sakai2, Suzuki, Tsuda2}) can also be derived from 
the cluster algebra corresponding to 
the same quiver $Q$ 
when $m=2$ and $n$ is even greater than two,
based on the present
framework  involving (\ref{eq:sd}).
\end{remark}

\section{Symplectic structure}
\label{sect:symp}

In this section we present a unified way to choose 
Darboux coordinates for the discrete dynamical systems arising from cluster algebras.
After stating the general result, we demonstrate the case of $q$-$P_{\rm VI}$ as a typical example.

Let $Q$ be a quiver with a vertex set $V=\{1,2,\ldots,N\}$ 
specified by a skew-symmetric integer matrix 
$B=(b_{ij})_{i,j=1}^N$,
i.e. $b_{ij} \in {\mathbb Z} $ and $b_{ij}=-b_{ji}$,
as its signed adjacency matrix.
After 
Gekhtman--Shapiro--Vainshtein \cite{GSV},
we define a Poisson bracket $\{ \cdot , \cdot  \}$
over the field 
${\mathbb Q}(y_1,y_2,\ldots,y_N)$ of rational functions
by
$\{ y_i, y_j\} = b_{ij} y_iy_j$,
which is compatible with any mutation $\mu_k$ 
in the sense that 
$\{ \mu_k(y_i), \mu_k(y_j)\} = \mu_k({b_{ij}})  \mu_k(y_i) \mu_k(y_j)$.
In particular, any element $w \in G_Q$ preserves the Poisson bracket $\{ \cdot , \cdot  \}$
since $w(B)=B$.
Fix the notation of multi-index
as $y^{\boldsymbol m}=\prod_{i=1}^N {y_i}^{m_i}$ and ${\boldsymbol m}={}^{\rm T}(m_1,m_2,\ldots,m_N)$.
By applying Leibniz's rule, 
we can verify the formula
\begin{equation}  \label{eq:poisson}
\{ y^{\boldsymbol m},y^{\boldsymbol n}\} =
\left({}^{\rm T}{\boldsymbol m} B {\boldsymbol n} \right) y^{{\boldsymbol m}+{\boldsymbol n}}
\end{equation}
for any ${\boldsymbol m}, {\boldsymbol n} \in {\mathbb Z}^N$.
Therefore, the Laurent monomial
$y^{\boldsymbol m}$ is a Casimir function 
with respect to
the Poisson bracket $\{\cdot,\cdot \}$
if and only if ${\boldsymbol m} \in \ker B$.

In general,
the mutation rule (\ref{eq:mut_y}) is rewritten as 
\begin{equation} \label{eq:mut_y_2}
\mu_k(y^{\boldsymbol m})= y^{A_k {\boldsymbol m}} (1+y_k)^{-(B{\boldsymbol m})_k}
\end{equation}
for any ${\boldsymbol m} \in {\mathbb Z}^N$,
where  $A_k \in GL_N(\mathbb Z)$ is the unimodular matrix defined in {\rm(\ref{eq:Ak})} 
and the symbol $(B{\boldsymbol m})_k$ denotes the $k$th component of the vector 
$B{\boldsymbol m} \in {\mathbb Z}^N$.
The next lemma implies that
the set of Casimir functions is closed under arbitrary mutations;
thus, Casimir functions can be regarded as parameters of the discrete dynamical system.

\begin{lemma}  \label{lemma:casimir}
If ${\boldsymbol m} \in \ker B \cap {\mathbb Z}^N$ then  
\[\mu_k (y^{\boldsymbol m})=y^{A_k{\boldsymbol m} } 
\quad
\text{and} \quad
A_k{\boldsymbol m} \in \ker \mu_k(B) \cap {\mathbb Z}^N. \]
\end{lemma}

\pf
The former is obvious from (\ref{eq:mut_y_2}).
The latter is a consequence of ${A_k}^2={\rm id}$
and
$\mu_k(B)={}^{\rm T}A_k B A_k$.
Recall (\ref{eq:mut_B}) and (\ref{eq:Ak}).
 \qed

\begin{lemma}[see {\cite[Theorems IV.1 and IV.2]{Newman}}]
\label{lemma:skewNF}
Let $B=(b_{ij})_{i,j=1}^N$ be a skew-symmetric integer matrix of rank $2 \ell$.
Then there exists a unimodular matrix  $  U=(u_{ij})_{i,j=1}^N \in GL_N(\mathbb Z)$
such that
\begin{equation}
\label{eq:skewNF}
 {}^{\rm T}U B U=
  \begin{pmatrix}  0& h_1 \\ -h_1 & 0\end{pmatrix} 
 \oplus   \begin{pmatrix}  0& h_2 \\ -h_2 & 0\end{pmatrix}\oplus \cdots \oplus
 \begin{pmatrix}  0& h_\ell \\ -h_\ell & 0\end{pmatrix} 
 \oplus O_{N-2\ell},
 \end{equation}
where 
the positive integers $h_1,h_2,\ldots,h_\ell$ satisfy $h_i | h_{i+1}$ and are uniquely determined by $B$.
\end{lemma}

This classical result about the normal form of an integer matrix is 
crucial to finding Darboux coordinates 
as seen below.
Write the above unimodular matrix as
\[
U= \left( {\boldsymbol u}_1,  \ldots, {\boldsymbol u}_{2\ell},   {\boldsymbol u}_{2\ell+1}, 
\ldots,   {\boldsymbol u}_N  \right)  \in GL_N(\mathbb Z)
\]
by arranging column vectors 
${\boldsymbol u}_j={}^{\rm T}(u_{1,j},u_{2,j},\ldots,u_{N,j})$
in a row.
We introduce  the  $2\ell$ variables
\[
f_i=y^{{\boldsymbol u}_{2i-1}}, \quad
g_i= y^{{\boldsymbol u}_{2i}} \quad (1 \leq i \leq  \ell)
\]
whose exponents are the first $2 \ell$ columns of $U$.
Combining (\ref{eq:poisson}) and  (\ref{eq:skewNF}),
we can readily verify
that
\begin{equation} \label{eq:ccr}
\{f_i,f_j\} =\{g_i,g_j\}=0 \quad \text{and}
\quad
\{f_i,g_j\} =h_i \delta_{ij} f_ig_j;
\end{equation}
these relations are preserved by any element of $G_Q$.
We also introduce the $N-2\ell$ variables
\[
\kappa_i=y^{ {\boldsymbol u}_{2\ell+i} } \quad (1 \leq i \leq  N-2\ell)
\]
whose exponents are the latter $N-2\ell$ columns of $U$;
thus, each of them is a Casimir function 
and plays the role of a parameter
in view of ${\boldsymbol u}_{2\ell+i} \in\ker B$ 
and Lemma~\ref{lemma:casimir}.
The unimodularity of $U$ asserts that 
the
variables $f_i$, $g_i$ and $\kappa_i$ are Laurent monomials in $y$-variables and vice versa.
This means that the birationality is preserved by this change of variables.

Summarizing above, we are led to the following theorem.

\begin{thm} \label{thm:symp}
Let ${\mathbb K}({\boldsymbol f},{\boldsymbol g})={\mathbb K}(f_1, \ldots,f_\ell,g_1,\ldots,g_\ell)$
be the field of rational functions of $f_i$ and $g_i$ $ (1 \leq i \leq  \ell)$
whose coefficient field is ${\mathbb K}={\mathbb Q}(\kappa_1, \ldots, \kappa_{N-2 \ell})$.
Then the action of $G_Q$ on ${\mathbb K}({\boldsymbol f},{\boldsymbol g})$ is birational and preserves the $2$-form
\[
\omega=
\sum_{i=1}^\ell
\frac{1}{h_i}
\frac{{\rm d}f_i \wedge {\rm d}g_i}{f_ig_i}
\]
except for signs.
\end{thm}

Assume that $h_i\equiv1$ $(1 \leq i \leq \ell)$ for simplicity.
Then $q_i=\log f_i$ and $p_i=\log g_i$ $(1 \leq i \leq \ell)$  
are Darboux coordinates for the symplectic $2$-form
$ \omega=\sum_{i=1}^\ell {\rm d} q_i \wedge {\rm d}p_i$
and the action of $G_Q$ gives rise to canonical transformations.
Note that 
in many concrete examples
relevant to the $q$-Painlev\'e equations and their higher-order extensions,
$h_i\equiv1$  occurs.

\begin{remark}\rm
For a special case where
the shape of $Q$ is invariant under a single mutation,
a similar statement of Theorem~\ref{thm:symp} was proved by 
Fordy--Hone \cite[Theorem~2.6]{FH}.
\end{remark}

Since $B$ is skew-symmetric,
it holds that $ \ker B \perp {\rm im \, } B$.
Therefore
we may
modify
$U$ by elementary column operations so that the first $2 \ell$  columns belong to ${\rm im \, } B$
as follows:
\begin{equation} \label{eq:tildeU}
\widetilde{U}=\left( \tilde{\boldsymbol u}_1, \ldots,   \tilde{\boldsymbol u}_{2 \ell},  {\boldsymbol u}_{2\ell+1},  
\ldots,   {\boldsymbol u}_N  \right)
=U\left(
\begin{array}{c|c}
I_{2\ell}&O\\
\hline
(r_{ij} )_{ \begin{subarray}{c} 
2 \ell +1 \leq i \leq N
\\ 
1  \leq j \leq 2 \ell
 \end{subarray}}&I_{N-2\ell}
\end{array}
\right),
\end{equation}
where $I_n= (\delta_{ij})_{i,j=1}^n$ is the identity matrix of size $n$
and $r_{ij} \in {\mathbb Q}$ are the rational numbers uniquely determined
by the condition
\[
\tilde{\boldsymbol u}_j={\boldsymbol u}_j+ \sum_{i=2 \ell +1}^N r_{ij}  {\boldsymbol u}_i
\in
{\rm im \, } B
\quad (1  \leq j \leq 2 \ell).
\]
Let us {\it redefine} the $2 \ell$ variables $f_i$ and $g_i$ by
\[
f_i=y^{\tilde{\boldsymbol u}_{2i-1}}, \quad
g_i= y^{\tilde{\boldsymbol u}_{2i}} \quad (1 \leq i \leq  \ell).
\]
Since 
${}^{\rm T}\widetilde{U} B \widetilde{U}={}^{\rm T}U B U$ holds, 
(\ref{eq:ccr}) is still valid.
Consequently, 
Theorem~\ref{thm:symp} 
holds true 
for these 
newly defined dynamical variables $f_i$ and $g_i$ 
with
the coefficient field 
${\mathbb K}={\mathbb Q}(\kappa_1, \ldots, \kappa_{N-2 \ell})$
replaced
with that generated by suitable fractional powers of $\kappa_i$. 
If a permutation $\sigma$ of vertices keeps $Q$ invariant, i.e. $\sigma(B)=B$,
then $\sigma$ acts trivially on $f_i$ and $g_i$ because $\tilde{\boldsymbol u}_j \in
{\rm im \, } B$ for $1  \leq j \leq 2 \ell$.

Finally, we shall demonstrate how to choose Darboux coordinates for 
the case of $q$-$P_{\rm VI}$:
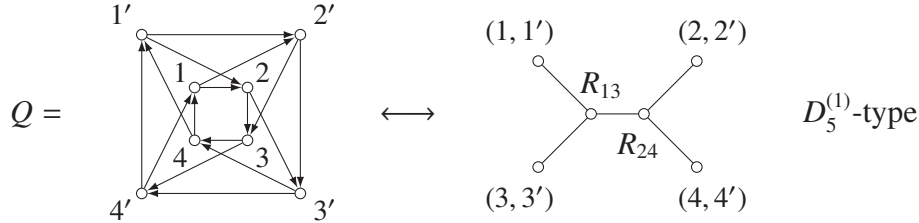
\begin{center}
\begin{picture}(300,80)
\put(0,37){$Q=$}
\put(70,50){\circle{4}}  
\put(62,52){\small$1$} 
\put(72,50){\vector(1,0){16}}
\put(90,50){\circle{4}}  
\put(93,52){\small$2$} 
\put(70,30){\circle{4}}
\put(62,20){\small$4$} 
\put(88,30){\vector(-1,0){16}}
\put(90,30){\circle{4}}  
\put(93,20){\small$3$} 
\put(70,32){\vector(0,1){16}}
\put(90,48){\vector(0,-1){16}}
\put(38,74){\small$1'$} 
\put(50,70){\circle{4}}  
\put(69,32){\vector(-1,2){18}}
\put(52,69){\vector(2,-1){36}}

\put(115,74){\small$2'$} 
\put(110,70){\circle{4}}  
\put(109,68){\vector(-1,-2){18}}
\put(72,51){\vector(2,1){36}}

\put(52,70){\vector(1,0){56}}  

\put(38,0){\small$4'$} 
\put(50,10){\circle{4}}  
\put(51,12){\vector(1,2){18}}
\put(88,29){\vector(-2,-1){36}}

\put(115,0){\small$3'$} 
\put(110,10){\circle{4}}  
\put(91,48){\vector(1,-2){18}}
\put(108,11){\vector(-2,1){36}}

\put(108,10){\vector(-1,0){56}}  
\put(50,12){\vector(0,1){56}}
\put(110,68){\vector(0,-1){56}}

\put(140,37){$\longleftrightarrow$}

\put(220,40){\circle{4}}  
\put(216,48){$R_{13}$} 
\put(222,40){\line(1,0){16}}
\put(240,40){\circle{4}}  
\put(230,25){$R_{24}$} 

\put(180,68){\small$(1,1')$} 
\put(200,60){\circle{4}} 
\put(201.5,58.5){\line(1,-1){17}}

\put(180,5){\small$(3,3')$} 
\put(200,20){\circle{4}} 
\put(201.5,21.5){\line(1,1){17}}

\put(253,68){\small$(2,2')$} 
\put(260,60){\circle{4}} 
\put(241.5,41.5){\line(1,1){17}}

\put(253,5){\small$(4,4')$} 
\put(260,20){\circle{4}} 
\put(241.5,38.5){\line(1,-1){17}}

\put(300,37){$D_5^{(1)}$-type}
\end{picture}
\end{center}
Recall Section~\ref{subsec:qp6}.
The signed adjacency matrix  is 
an $8 \times 8$ skew-symmetric integer matrix
\[
B=(b_{ij})_{i,j=1,2,3,4,1',2',3',4'}
=
\begin{pmatrix}
0&1&0&-1& 0&1&0&-1 \\
-1&0&1&0&-1& 0&1&0 \\
0&-1&0&1&0&-1& 0&1 \\
1&0&-1&0&1&0&-1& 0 \\
0&1&0&-1& 0&1&0&-1 \\
-1&0&1&0&-1& 0&1&0 \\
0&-1&0&1&0&-1& 0&1 \\
1&0&-1&0&1&0&-1& 0
\end{pmatrix}
\begin{array}{l}
1 \\ 2 \\  3\\   4\\ 1' \\ 2' \\ 3' \\ 4'
\end{array}
\]
and  ${\rm rank \, } B=2$.
The matrix
$B$ is congruent to the following normal form (see Lemma~\ref{lemma:skewNF}):
\[{}^{\rm T}U B U= \begin{pmatrix}  0&1 \\-1 & 0\end{pmatrix}
\oplus O_6,
\]
where the unimodular matrix $U$ can be taken as
\[
U=\left(
\begin{array}{cc|cccccc}
1&0&1&0&-1&0&0&0
\\
0&1&0&1&0&-1&0&0
\\
0&0&1&0&0&0&-1&0
\\
0&0&0&1&0&0&0&-1
\\
0&0&0&0&1&0&0&0
\\
0&0&0&0&0&1&0&0
\\
0&0&0&0&0&0&1&0
\\
0&0&0&0&0&0&0&1
\end{array}
\right)
\in GL_8({\mathbb Z}).
\]
We modify
$U$ so that the first two columns belong to ${\rm im \, } B$
as
\[
\widetilde{U}=
U
\left(
\begin{array}{c|c}
I_2 & 
\\
\hline
\begin{matrix}
-1/2 & 0
\\
0 & -1/2
\\
1/4&0
\\
0&1/4
\\
-1/4&0
\\
0&-1/4
\end{matrix}
& I_6
\end{array}
\right)
=
\left(
\begin{array}{cc|cccccc}
1/4&  0  &1&0&-1&0&0&0
\\
0&1/4 &   0&1&0&-1&0&0
\\
-1/4&0 &  1&0&0&0&-1&0
\\
0&-1/4 &0&1&0&0&0&-1 
\\
1/4&0 & 0&0&1&0&0&0
\\
0&1/4&0&0&0&1&0&0
\\
-1/4&0&0&0&0&0&1&0
\\
0&-1/4&0&0&0&0&0&1
\end{array}
\right).
\]
Although $\widetilde{U}=(\tilde{\boldsymbol u}_1, \tilde{\boldsymbol u}_2, {\boldsymbol u}_3, \ldots, {\boldsymbol u}_8)$ is no longer an integer matrix, 
it still holds that
\[{}^{\rm T}\widetilde{U} B \widetilde{U}= \begin{pmatrix}  0&1 \\-1 & 0\end{pmatrix}
\oplus O_6.
\]
Define the dynamical variables $f$ and $g$ by
\[
f= y^{\tilde{\boldsymbol u}_1}=
\left( \frac{y_1y_{1'}}{y_3 y_{3'}} \right)^{1/4}, \quad
g= y^{\tilde{\boldsymbol u}_2}=\left( \frac{y_2y_{2'}}{y_4 y_{4'}} \right)^{1/4}
\]
and the  {\it multiplicative root variables}
$a_i$ $( 0 \leq i \leq 5)$ 
by
\begin{align*}
a_2&=y^{{\boldsymbol u}_3}=y_1y_3, \quad 
a_3=y^{{\boldsymbol u}_4}=y_2y_4,
\\
a_0&=y^{{\boldsymbol u}_5}=\frac{y_{1'}}{y_1}, \quad 
a_5=y^{{\boldsymbol u}_6}=\frac{y_{2'}}{y_2}, \quad
a_1=y^{{\boldsymbol u}_7}=\frac{y_{3'}}{y_3}, \quad
a_4=y^{{\boldsymbol u}_8}=\frac{y_{4'}}{y_4}.
\end{align*}
Then the birational realization (\ref{eq:D5}) and (\ref{eq:D5_auto})
of $\widetilde{W}(D_5^{(1)})=\langle s_i \ (0 \leq i \leq 5) \rangle \rtimes \langle
\sigma_1, \sigma_2 \rangle$
is rewritten into the following form:
\begin{equation}   \label{eq:D5onfg}
\begin{aligned}
&s_i(a_j)=a_j {a_i}^{-c_{ij}}, \\
&\frac{s_2(g)}{g}=
 {a_2}^{-1/2} \frac
 {f+ {a_0}^{1/4}{a_1}^{-1/4}{a_2}^{1/2}}
 {f+ {a_0}^{1/4}{a_1}^{-1/4}{a_2}^{-1/2} },
\quad
\frac{s_3(f)}{f}= {a_3}^{1/2}
 \frac{g+ {a_3}^{-1/2}{a_4}^{-1/4}{a_5}^{1/4} }{g+ {a_3}^{1/2}{a_4}^{-1/4}{a_5}^{1/4}},
\\
&\sigma_1(a_{\{ 0,1,2,3,4,5\}})=\frac{1}{ a_{\{5,4,3,2,1,0 \}} },
\quad\sigma_1(f)=\frac{1}{g}, \quad \sigma_1(g)=\frac{1}{f}, 
\\
&\sigma_2(a_{\{0,1,2,3,4,5\}})=\frac{1}{a_{\{1,0,2,3,4,5 \}}},
\quad
 \sigma_2(g)= \frac{1}{g}
\end{aligned}
\end{equation}
with  $(c_{ij})_{0 \leq i,j \leq 5}$ 
being the Cartan matrix of type $D_5^{(1)}$.
The action (\ref{eq:D5onfg}) of $\widetilde{W}(D_5^{(1)})$ on ${\mathbb K}(f,g)$ is certainly birational, where the coefficient field is 
${\mathbb K}={\mathbb Q}(\{{a_i}^{1/4} \, |\, 0 \leq i \leq 5\})$.
Moreover, 
$\omega=(fg)^{-1}{\rm d} f \wedge {\rm d}g$
is invariant under the action of $W(D_5^{(1)})$,
i.e. $s_i(\omega)=\omega$ $(0 \leq i \leq 5)$, 
and $\sigma_j(\omega)=-\omega$ $(j=1,2)$.
Therefore,
the pair of variables $(\log f,\log g)$ provides Darboux coordinates for the symplectic $2$-form $\omega$.
Let $q$ denote the product of all the $y$-variables, which 
amounts to the multiplicative null root:
$q=y_1y_2y_3y_4y_{1'}y_{2'}y_{3'}y_{4'}
=a_0a_1{a_2}^2{a_3}^2a_4a_5$.
A translation $T=(\sigma_1 \sigma_2 s_2 s_0s_1s_2)^2 \in \widetilde{W}(D_5^{(1)})$ gives rise to the non-autonomous system 
\begin{equation} \label{eq:qp6}
\begin{aligned}
&f[n+1]f[n]
\\
&\quad =\frac{1}{q^na_3{a_4}^{1/2}{a_5}^{1/2}}\frac{(g[n]+q^{n/2}{a_3}^{1/2}{a_4}^{-1/4}{a_5}^{1/4})(g[n]+q^{n/2}{a_3}^{1/2}{a_4}^{3/4}{a_5}^{1/4})}{(g[n]+q^{-n/2}{a_3}^{-1/2}{a_4}^{-1/4}{a_5}^{1/4})(g[n]+q^{-n/2}{a_3}^{-1/2}{a_4}^{-1/4}{a_5}^{-3/4})},
\\
&g[n]g[n-1]
\\
& \quad
=\frac{{a_0}^{1/2}{a_1}^{1/2}a_2}{q^{n}}\frac{(f[n]+q^{n/2}{a_0}^{1/4}{a_1}^{-1/4}{a_2}^{-1/2})(f[n]+q^{n/2}{a_0}^{-3/4}{a_1}^{-1/4}{a_2}^{-1/2})}{(f[n]+q^{-n/2}{a_0}^{1/4}{a_1}^{-1/4}{a_2}^{1/2})(f[n]+q^{-n/2}{a_0}^{1/4}{a_1}^{3/4}{a_2}^{1/2})}
\end{aligned}
\end{equation}
of $q$-difference equations
for unknowns $f[n]=T^n(f)$ and $g[n]=T^n(g)$,
which is exactly 
the sixth $q$-Painlev\'e equation ($q$-$P_{\rm VI}$); 
cf. \cite{JS, Sakai1}.

\small
\paragraph{\it Acknowledgement.}
The authors are deeply grateful to
Rei Inoue, Yuma Mizuno, Takao Suzuki and Yuji Terashima
for 
invaluable discussions and comments on various aspects of cluster algebras.
This work was partly conducted during the program
``Aspects of Combinatorial Representation Theory" (2018)
in the Research Institute for Mathematical Sciences, Kyoto University.
It was also supported  
by a grant-in-aid from the Japan Society for the Promotion of Science (Grant Number JP17K05270).



\ \\
\begin{quote}
Tetsu Masuda \\
Department of Mathematical Sciences, Aoyama Gakuin University, 
Kanagawa 252-5258, Japan.
\\
e-mail: masuda@math.aoyama.ac.jp
\\
\\
Naoto Okubo \\
Department of Mathematical Sciences, Aoyama Gakuin University, 
Kanagawa 252-5258, Japan.
\\
e-mail: okubo@math.aoyama.ac.jp
\\
\\
Teruhisa Tsuda \\
Department of Mathematical Sciences, Aoyama Gakuin University, 
Kanagawa 252-5258, Japan.
\\
e-mail: tudateru@math.aoyama.ac.jp
\end{quote}


\small
\begin{thebibliography}{99}
%

\bibitem{BGM}
 Bershtein, M.,
 Gavrylenko, P.,
  Marshakov, A.: 
 Cluster integrable systems, $q$-Painlev\'e equations and their quantization.
 J. High Energy Phys. {\bf 2018}, 077 (33pp) (2018)
 
%
\bibitem{Bucher}
Bucher, E.:
Maximal green sequences for cluster algebras associated to the $n$-torus. 
arXiv:1412.3713

\bibitem{DO}
 Dolgachev, I., Ortland, D.:
 Point sets in projective spaces and theta functions.
 Ast\'erisque {\bf 165} (1988)

\bibitem{FZ}
 Fomin, S., Zelevinsky, A.:
 Cluster algebras. IV. Coefficients.
 Compos. Math. {\bf 143}, 112--164 (2007) 
 
\bibitem{FH}
Fordy,  A.P.,  Hone, A.:
Discrete integrable systems and Poisson algebras from cluster maps.
Comm. Math. Phys. {\bf 325}, 
527--584 (2014) 

%
\bibitem{GSV}
Gekhtman, M., Shapiro, M., Vainshtein, A.:
Cluster Algebras and Poisson Geometry.
American Mathematical Society, Providence (2010)
%
\bibitem{GS}
 Goncharov, A., Shen, L.: 
 Donaldson-Thomas transformations of moduli spaces of $G$-local systems.
 Adv. in Math. {\bf 327}, 225--348 (2018) 
%

\bibitem{HI}
Hone, A., Inoue, R.: 
Discrete Painlev\'e equations from Y-systems. 
J. Phys. A: Math. Theor. {\bf 47}, 
474007 (26pp) (2014) 

\bibitem{IIO}
Inoue, R.,  Ishibashi, T., 
Oya, H.:
Cluster realization of Weyl groups and higher 
Teichm\"uller theory.
Selecta Math. (N.S.) {\bf 27}, 37 (84pp) (2019)
%

\bibitem{ILP}
Inoue, R., Lam, T., Pylyavskyy, R.:
On the cluster nature and quantization of geometric $R$-matrices.
 Publ. Res. Inst. Math. Sci. {\bf 55}, 25--78 (2019)
%

\bibitem{JS}
Jimbo, M., Sakai, H.: A $q$-analog of the sixth Painlev\'e equation. 
Lett. Math. Phys. {\bf 38}, 145--154 (1996)

\bibitem{KMNOY}
Kajiwara, K.,   Masuda,  T.,  Noumi, M., Ohta, Y.,  Yamada, Y.:
Point configurations, Cremona transformations and the elliptic difference Painlev\'e equation.
S\'emin. Congr. {\bf 14}, 169--198 (2006)


\bibitem{KNY}
  Kajiwara, K.,  Noumi, M.,   Yamada, Y.: 
 Discrete dynamical systems with $W(A^{(1)}_{m-1}\times A^{(1)}_{n-1})$ symmetry.
 Lett. Math. Phys. {\bf 60},
 211--219 (2002)

\bibitem{Kirillov}
Kirillov, A.N.:
Introduction to tropical combinatorics.
In: Kirillov, A.N., Liskova, N. (eds.),
Physics and Combinatorics 2000,
pp. 82--150.
World Scientific, Singapore (2001)

\bibitem{Looijenga}
Looijenga, E.:
Rational surfaces with an anticanonical cycle. 
Ann. of Math. {\bf 114}, 
267--322 (1981)

\bibitem{Masuda}
Masuda, T.:
A $q$-analogue of the higher order Painlev\'e type equations with the affine Weyl group symmetry of type $D$.
Funkcial. Ekvac. {\bf 58}, 
405--430 (2015) 

\bibitem{MMOTT}
Masuda, T.,   Mizuno, Y., Okubo, N., Terashima, Y., Tsuda, T.:
Birational Weyl group actions and $q$-Painlev\'e equations via mutation combinatorics in cluster algebras II: $\tau$-Function formalism.
(a tentative title)
(in preparation)

\bibitem{MOT}
Masuda, T., Okubo, N., Tsuda, T.:
Cluster algebras and higher order generalizations of the $q$-Painlev\'e equations
of type $A_7^{(1)}$ and $A_6^{(1)}$.
RIMS Kokyuroku Bessatsu {\bf B87},
149--163  (2021)

\bibitem{Mizuno}
Mizuno, Y.:
$q$-Painlev\'e equations on cluster Poisson varieties via toric geometry.
Selecta Math. (N.S.) {\bf 30},  19 (37pp) (2024)

\bibitem{Newman}
Newman, M.: {\it Integral matrices}. Academic Press.1972

\bibitem{ORG}
Ohta, Y., Ramani, A., Grammaticos, B.: An affine Weyl group approach to the eight-parameter discrete Painlev\'e equation. 
J. Phys. A: Math. Gen. {\bf 34}, 
10523--10532 (2001)

\bibitem{Okubo1}
 Okubo, N.:
 Discrete integrable systems and cluster algebras.
 RIMS Kokyuroku Bessatsu {\bf B41},
 25--41 (2013) 
%
\bibitem{Okubo2}
Okubo, N.:
 Bilinear equations and $q$-discrete Painlev\'e equations satisfied by variables and coefficients in cluster algebras.
 J. Phys. A: Math. Theor. {\bf 48},
 355201  (2015) (25pp)
 
\bibitem{OS}
 Okubo, N.,  Suzuki, T.:
 Generalized $q$-Painlev\'e VI systems of type $(A_{2n+1}+A_1+A_1)^{(1)}$ arising from cluster algebra.
 Int. Math. Res. Not. {\bf 2022},
 6561--6607 (2022)


\bibitem{Sakai1}
Sakai, H.:
Rational surfaces associated with affine root systems 
and geometry of the Painlev\'e equations. 
Comm. Math. Phys. {\bf 220},
165--229 (2001)

\bibitem{Sakai2}
Sakai, H.:
A $q$-analog of the Garnier system. 
Funkcial. Ekvac. {\bf 48}, 
273--297
(2005)

\bibitem{Suzuki}
Suzuki, T.:
A $q$-analogue of the Drinfeld-Sokolov hierarchy of type $A$ and
$q$-Painlev\'e system.
AMS Contemp. Math. {\bf 651},
25--38 (2015) 

\bibitem{Takenawa}
Takenawa, T.:
Discrete dynamical systems associated with the configuration space of 8 points in ${\mathbb P}^3({\mathbb C})$.
Comm. Math. Phys. {\bf 246}, 
19--42 (2004)

%
\bibitem{TTMS} Tokihiro, T., Takahashi, D., Matsukidaira, J.,  Satsuma, J.:
From soliton equations to integrable cellular automata through a limiting procedure.  Phys. Rev. Lett. {\bf 76},  3247--3250 (1996)
%



\bibitem{Tsuda1}
Tsuda, T.:
 Tropical Weyl group action via point configurations and $\tau$-functions of the $q$-Painlev\'e equations. 
 Lett. Math. Phys. {\bf 77},
 21--30 (2006)

\bibitem{Tsuda2}
Tsuda, T.:
On an integrable system of $q$-difference equations satisfied by the universal characters: its Lax formalism and an application to $q$-Painlev\'e equations.
Comm. Math. Phys. {\bf 293},
 347--359 (2010) 

\bibitem{TT}
 Tsuda, T., Takenawa, T.:
 Tropical representation of Weyl groups associated with certain rational varieties.
 Adv. Math. {\bf 221},
 936--954  (2009) 
%
\bibitem{Yamada}
Yamada, Y.:
A birational representation of Weyl group, combinatorial
$R$-matrix and discrete Toda equation.
In: Kirillov, A.N., Liskova, N. (eds.),
Physics and Combinatorics 2000,
pp. 305--319.
World Scientific, Singapore (2001)
%
\end{thebibliography}
\end{document}